\documentclass[graybox, nosecnum]{svmult}

\usepackage{mathptmx}       
\usepackage{helvet}         
\usepackage{courier}        
\usepackage{type1cm}        
\usepackage{makeidx}         
\usepackage{graphicx}        
\usepackage{siunitx}
\usepackage{chemformula}
\usepackage{tikz}
\usetikzlibrary{circuits.ee.IEC}
\usepackage{array}
\usepackage{multicol}        
\usepackage[bottom]{footmisc}
\usepackage{hyperref}        
\usepackage{soul}            
\hypersetup{colorlinks=true,urlcolor=blue}
\usepackage[square,numbers]{natbib}
\makeindex             

\newcommand\Ground{%
\mathbin{\text{\begin{tikzpicture}[circuit ee IEC,yscale=0.6,xscale=0.5]
\draw (0,2ex) to (0,0) node[ground,rotate=-90,xshift=.65ex] {};
\end{tikzpicture}}}%
}

\begin{document}

\title*{Technologies for advanced X-ray mirror fabrication}
\author{Carolyn Atkins \thanks{corresponding author}}
\institute{Carolyn Atkins \at STFC UK Astronomy Technology Centre, Blackford Hill, Edinburgh, UK\\ \email{carolyn.atkins@stfc.ac.uk}}
%
%
\maketitle
\abstract{X-ray mirror fabrication for astronomy is challenging; this is due to the Wolter I optical geometry and the tight tolerances on roughness and form error to enable accurate and efficient X-ray reflection. The performance of an X-ray mirror, and ultimately that of the telescope, is linked to the processes and technologies used to create it. The goal of this chapter is to provider the reader with an overview of the different technologies and processes used to create the mirrors for X-ray telescopes. The objective is to present this diverse field in the framework of the manufacturing methodologies (subtractive, formative, fabricative \& additive) and how these methodologies influence the telescope attributes (angular resolution and effective area). The emphasis is placed upon processes and technologies employed in recent X-ray space telescopes  and those that are being actively investigated for future missions such as \textit{Athena} and concepts such as \textit{Lynx}. Speculative processes and technologies relating to Industry 4.0 are introduced to imagine how X-ray mirror fabrication may develop in the future.}
\section{Keywords} 

X-ray mirrors, mirror fabrication, subtractive, formative, fabricative, additive, replication, active control.

\section{Introduction}
The goal of a telescope is to provide the astronomers with photons to analyse and the objective of a telescope is to deflect the photons from the field of view to the detector. The key component of a telescope is the primary mirror (or lens). The primary mirror must collect as many photons as possible and accurately deflect the photons to the focus or detector. A normal incidence telescope - e.g. \textit{Hubble Space Telescope}, angle of incidence $\theta_{i}=$ \ang{90;;} - achieves this by having a mirror as large as possible (\SI{2.4}{\metre} diameter) and polished as accurately as possible. The primary mirror for a grazing incidence telescope required for X-ray astronomy is based upon the Wolter I geometry, where photons hit the surface at very small angles of incidence ($\theta_{i}\approx$\ang{1;;}, or less). Therefore to fill the telescope aperture, to collect as many photons as possible, multiple grazing incidence mirrors are nested and as such, the `primary mirror' is made up from many X-ray mirrors. The challenge then becomes how to make the individual X-ray mirrors. 

Normal incidence mirrors can be made thick (a $\sim$ 6:1 - diameter to thickness ratio is often used) to ensure that the mirror surface is rigid; however, thick mirrors for an X-ray telescope reduce the telescope aperture (collecting area) because they are `seen' by the source edge-on and limit the number of mirrors that can be nested in the system. In contrast, thin mirrors block less of the telescope aperture and more mirrors can be nested, but the accuracy of the mirror surface is lower. 

There are multiple methods to create an X-ray mirror. The technology and process used depends upon the science objectives of the telescope/observatory - i.e. is imaging, number of photons, or both, required by the telescope. This chapter will explore the plethora of technologies and processes used in X-ray mirror fabrication, both in use today and under development for the future. To limit the scope of the chapter, only the technologies and processes that deliver $<$\ang{;1;} angular resolution (half power diameter; HPD) in the Wolter I configuration - i.e. \textit{Chandra X-ray Observatory} (\textit{Chandra}; \ang{;;0.5}) $\longrightarrow$ \textit{NuSTAR} (\ang{;;58}) - are described in detail; the rationale behind this down selection is to focus on those technologies that have the potential to deliver both high angular resolution and effective area in the future.

The technologies and processes are grouped by manufacture methodology: subtractive, formative, fabricative and additive, and we will see how these methodologies directly impact the science objectives. In addition, considerations such as, production volume, manufacture heritage and automation, are discussed where appropriate. This chapter is divided into two sections. First, the core background information is provided describing: the different manufacturing methodologies, how to evaluate the processes and technologies, and the fundamental terminology of X-ray mirrors used in this chapter. Second, the processes and technologies are presented in the appropriate manufacture methodology: subtractive, formative, fabricative and additive, respectively. Finally, bear in mind that one process and technology is not better than another; processes and technologies, particularly in X-ray mirrors, are developed to serve a given set of science objectives (and budget) and should be considered in that context. 

\section{X-ray mirror fabrication - fundamentals}

\subsection{Manufacturing methodologies}

\begin{figure}
    \centering
    \includegraphics[width = 7.5cm]{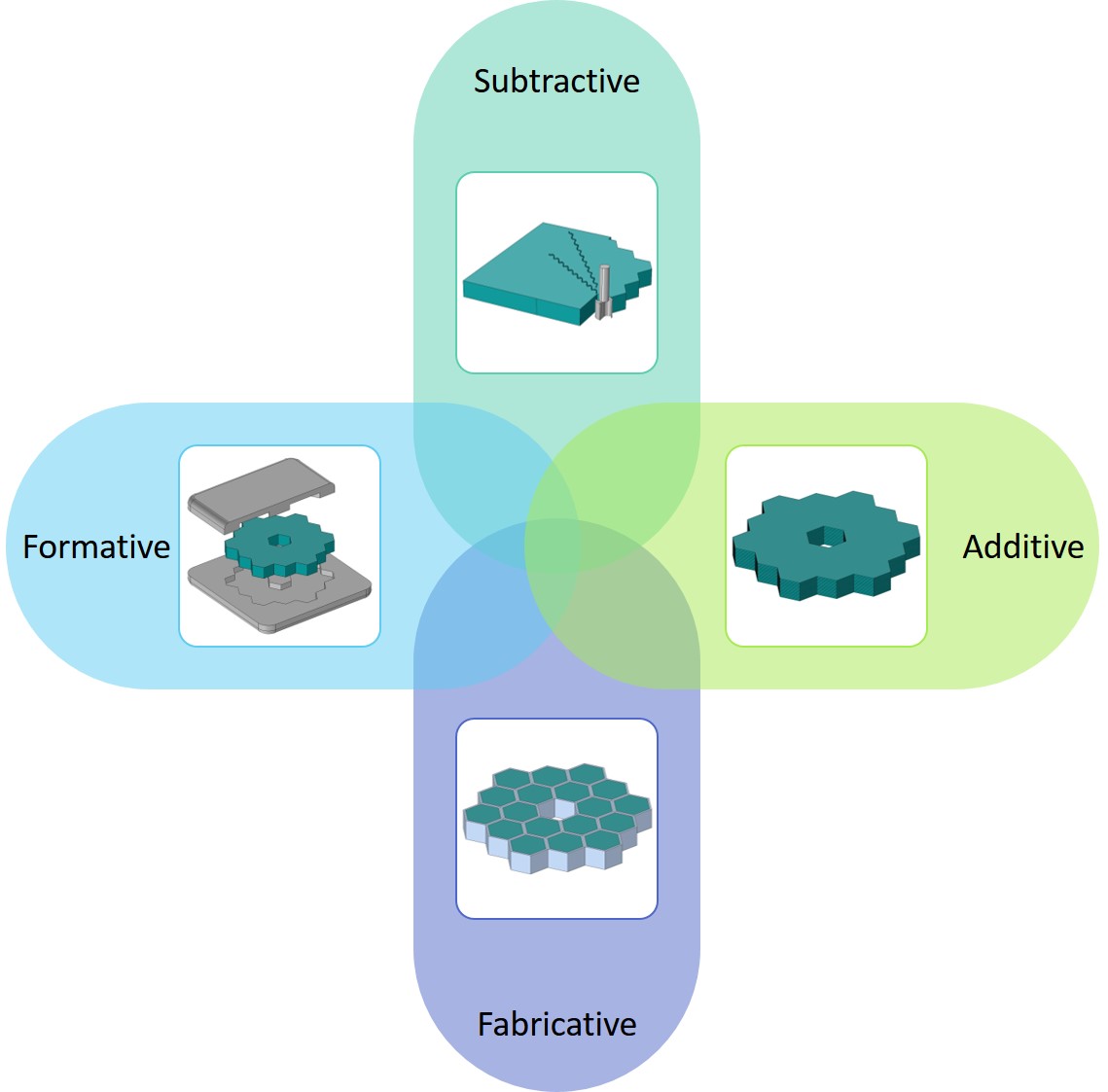}
    \caption{The interplay between the different manufacturing methodologies: subtractive, formative, fabricative and additive.}
    \label{fig:manu}
\end{figure}

Manufacturing is the process of converting a raw material into a functional part or component. Manufacturing methods describe the action of the process and fall into four general categories: subtractive (mill, drill \& lathe), formative (casting \& forging), fabricative (fixtures \& bonding) and additive (3D printing). Typically, one or more of the manufacturing methods will be utilised to create the functional part (Figure~\ref{fig:manu}); for example a wooden chair would have each of its individual parts subtractively machined from the bulk material and then the parts would be fabricatively combined via screws, glue, or joints, to create the functional component. In this chapter, \textit{technology} refers to the tool that is used to enable the \textit{process}: a lathe used for creating the chair legs is a technology, whereas the act removing the material is the process. Furthermore, in context of the chapter heading `\textit{Technologies for advanced X-ray mirror fabrication}', both the technology and the process can be considered \textit{advanced}.

\subsection{X-ray mirror manufacture and technology}

\begin{figure}
    \centering
    \includegraphics[width = 6.8cm]{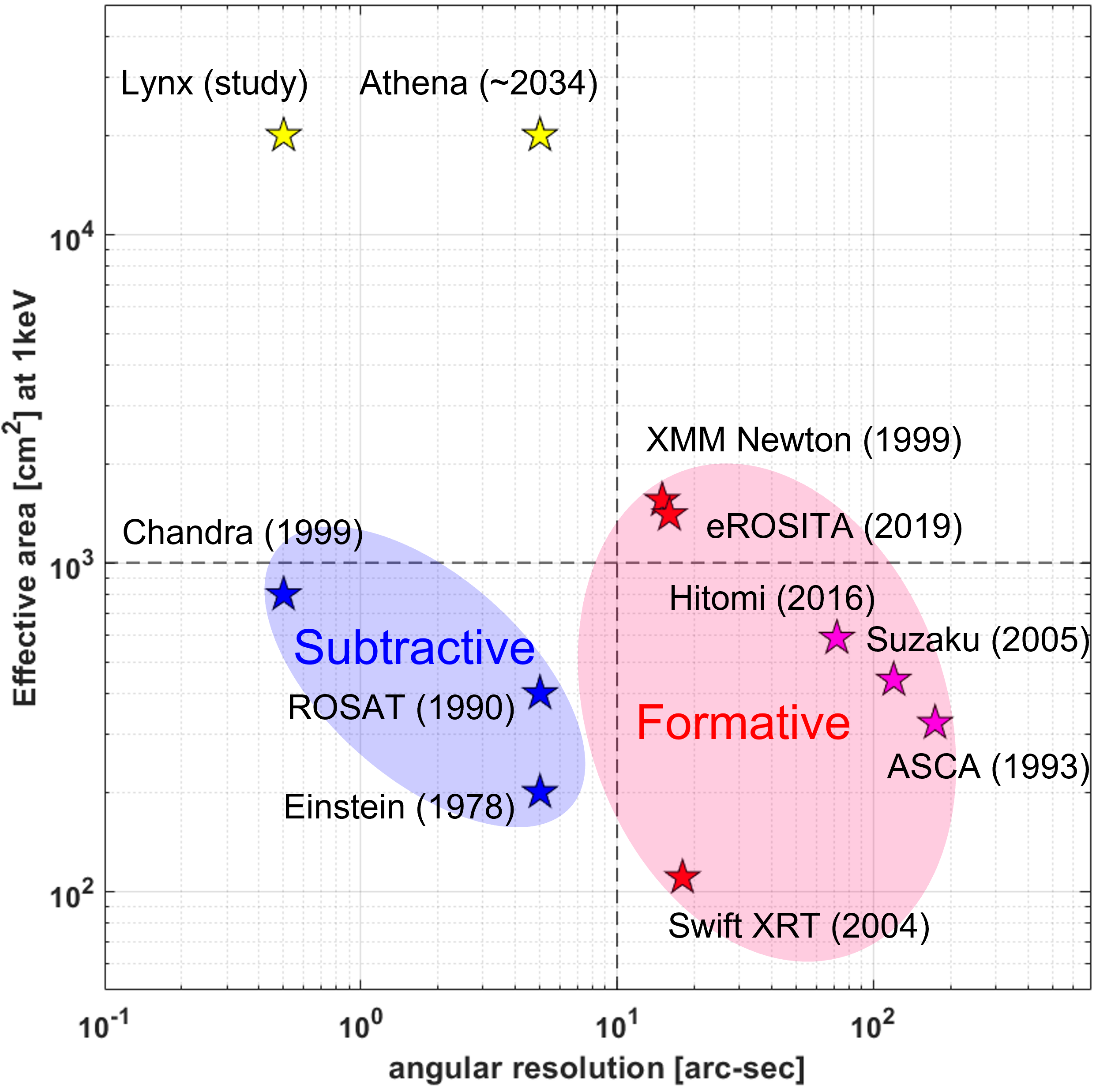}
    \caption{X-ray telescope missions employing a Wolter I optical configuration, plotted against angular resolution and effective area at \SI{1}{\kilo\electronvolt}, and grouped by primary manufacturing methodology.}
    \label{fig:xraymanu}
\end{figure}

To date, numerous technologies and processes have been used to create X-ray telescope mirrors. In the context of the methodologies, the \textit{Chandra} mirrors used grinding and polishing processes to convert cast Zerodur blanks into high precision X-ray mirrors, as such, the dominant methodology is subtractive. Figure~\ref{fig:xraymanu} presents the existing and historical X-ray telescope missions that utilise the Wolter I focussing geometry demonstrating the interplay between effective area and angular resolution at \SI{1}{\kilo\electronvolt}. The telescopes have been grouped in terms of primary manufacturing methodology and it is clear that the telescopes with the highest angular resolution (i.e. better resolving capability) are created via subtractive methodologies, whereas the telescopes with the largest effective area utilise formative methodologies.  

\begin{table}[]
    \centering
    \begin{tabular}{m{2cm}m{1cm}m{1cm}m{2.1cm}m{1.7cm}m{1.9cm}m{1.9cm}}
    Mission & Year & X-ray \newline range & Objective \newline (Ang. res.\ddag) &  $^{\star}$(Modules) \newline / Full shells & Material \newline (coating) & Production \newline (Manu.)\\[10pt]
         \hline
        $~$ \newline Einstein~\cite{Speybroeck1977} \newline $~$ & 1978 & Soft & Imaging \&\newline spectroscopy ($<$\ang{;;10}) & (1) /  4 & Fused quartz \newline (Ni) & Direct polish \newline (\textcolor{blue}{Subtractive}) \\
        $~$ \newline RT-4M \newline Salyut 7~\cite{MANDELSTAM1982} \newline $~$ & 1982 & Soft & Spectroscopy \newline ($<$\ang{;1;}) & (1) / 2 & Ni + epoxy \newline(Au + Ni) & Electroforming \newline (\textcolor{red}{Formative}) \\
       $~$ \newline EXOSAT~\cite{Laine1979} \newline $~$  & 1983 & Soft & Imaging \&\newline spectroscopy \newline ($<$\ang{;;20}) & (1) / 2 & Beryllium \newline + epoxy \newline(Au) & Replication \newline (\textcolor{red}{Formative}) \\
       $~$ \newline ROSAT~\cite{Aschenbach1988} \newline $~$ & 1990 & Soft & All-sky survey \newline ($<$\ang{;;10}) &(1) / 4 & Zerodur \newline(Au) & Direct polish \newline (\textcolor{blue}{Subtractive}) \\
       $~$ \newline ASCA~\cite{Serlemitsos1995} \newline $~$ & 1993 & Soft & Spectroscopy \newline  ($\sim$\ang{;3;}) &(4) / 120 & Al foil \newline + lacquer \newline(Au) & Replication \newline (\textcolor{red}{Formative}) \\
       $~$ \newline Chandra~\cite{WEISSKOPF2003v2} \newline $~$ & 1999 & Soft & Sub-arcsec \newline imaging \newline ($<$\ang{;;1}) & (1) / 4 & Zerodur \newline(Ir) & Direct polish \newline (\textcolor{blue}{Subtractive}) \\
       $~$ \newline XMM \newline Newton~\cite{deChambure1996} \newline $~$ & 1999 & Soft & Spectroscopy \newline ($<$\ang{;;20}) &(3) / 58 & Ni \newline(Au) & Electroforming \newline (\textcolor{red}{Formative})\\
       $~$ \newline Swift \newline XRT~\cite{Burrows2003} \newline $~$ & 2004 & Soft & $\gamma$-ray burst \newline locator \newline ($<$\ang{;;20}) &(1) / 12 & Ni \newline(Au) & Electroforming \newline (\textcolor{red}{Formative}) \\
       $~$ \newline Suzaku \newline XRT-I~\cite{Serlemitsos2007} \newline $~$ & 2005 & Soft & Spectroscopy \newline ($\sim$\ang{;2;}) & (4) / 175 & Al foil \newline + epoxy \newline(Au) & Replication \newline (\textcolor{red}{Formative}) \\
        $~$ \newline NuSTAR~\cite{Harrison2013} \newline $~$ & 2012 & Hard & Hard X-ray \newline astronomy \newline ($<$\ang{;1;}) & (2) / 133 & Borofloat \newline glass \newline(Multilayer) & Replication \newline (\textcolor{red}{Formative}) \\
        $~$ \newline Hitomi \newline SXT~\cite{Awaki14} \newline $~$ & 2016 & Soft & Calorimeter \newline spectroscopy \newline ($<$\ang{;2;}) & (2) / 203 & Al foil \newline + epoxy \newline(Au) & Replication \newline (\textcolor{red}{Formative}) \\
        $~$ \newline Hitomi \newline HXT~\cite{Awaki14} \newline $~$ & 2016 & Hard & Spectroscopy \newline ($<$\ang{;2;}) &(2) / 213 & Al foil \newline + epoxy \newline (Multilayer) & Replication \newline (\textcolor{red}{Formative}) \\
        $~$ \newline Spektr-RG \newline eROSITA~\cite{Friedrich2012} \newline $~$ & 2019 & Soft & All-sky survey \newline ($<$\ang{;;20}) &(7) / 54 & Ni \newline(Au) & Electroforming \newline (\textcolor{red}{Formative}) \\
        $~$ \newline Spektr-RG \newline ART-XC~\cite{Pavlinsky2018} \newline $~$ & 2019 & Hard & All-sky survey \newline ($<$\ang{;;40}) & (7) / 28 & NiCo \newline(Ir) & Electroforming \newline (\textcolor{red}{Formative}) \\
        $~$ \newline IXPE~\cite{Ramsey2019} \newline $~$ & 2021 & Soft & Polarimetry \newline ($<$\ang{;;40}) & (3) / 24 & NiCo \newline(-) & Electroforming \newline (\textcolor{red}{Formative}) \\
        \hline
        \multicolumn{4}{l}{\ddag - Grouped angular spatial resolution (HPD)} & \multicolumn{3}{l}{$^{\star}$ - (\# modules) / \# of $360^{\circ}$ shells per module}\\
    \end{tabular}
    \caption{Space-based X-ray telescopes (Wolter I configuration), ranked by year of launch and with emphasis on fabrication methods and materials. HPDs are approximated into groups for clarity.}
    \label{tab:xraytele}
\end{table}

\subsubsection{Angular resolution versus effective area}

The science objectives (and budget) of the telescope defines the required angular resolution and effective area; however, it is the manufacturing methodology that delivers the requirements. Table~\ref{tab:xraytele} highlights how the manufacturing methodology affects the angular resolution and effective area of the telescope; subtractive methodologies lead to high angular resolution and low effective area, whereas formative methodologies favour low angular resolution and large effective area. How to combine high angular resolution and large effective area into one system is the key challenge in the field of astronomical X-ray optics.      

Angular resolution provides a description of how accurately the mirrors reflect the photons to the focal/detector plane. A low numerical value (\textit{Chandra}: \ang{;;0.5}) is considered high angular resolution, whereas a high numerical value (\textit{NuSTAR}: \ang{;;58}) is considered a low angular resolution. The angular resolution is calculated from the Point Spread Function (PSF) of the focussed, on-axis image. For X-ray telescopes, angular resolution is commonly defined by the HPD, also termed Half Energy Width (HEW), which is the angular diameter that contains half of the flux of the PSF. This metric is different to the Full Width Half Maximum (FWHM), which defines the angular diameter at half the maximum flux of the PSF. 

Effective area is the area of the mirror surface that is available for the photons to interact with, which is not equivalent to the total area of the mirror surface. X-ray reflection occurs at grazing incidence, as X-rays at normal incidence are either absorbed or transmitted by the mirror surface. Therefore, the requirement of grazing incidence results in only a projection of the mirror surface being available to the X-rays for reflection - this surface projection provides the geometric component of the effective area. To calculate the effective area the geometric projection is multiplied by the X-ray reflectivity at a given $\theta_{i}$ and energy for a given mirror coating.     

\subsubsection{Production drivers for future X-ray telescopes}
A key driving factor in the development of technology for X-ray telescope mirrors is the production chain for the mirrors - how many mirrors are required and, ultimately, the time and budget. X-ray telescopes are not consumer products; X-ray observatories are typically operated by a space agency, or collaboration of space agencies, for a community of scientists. Therefore, in terms of product count, only one observatory is required. However, the style of the telescope (\# mirror modules \& \# mirror shells) within the observatory is variable, for example \textit{Chandra} contains only 1 module with 4 mirror shells, whereas \textit{eROSITA} contains 7 modules each with 54 mirror shells. Previous observatories have utilised a unit or batch production approach to mirror fabrication; however, future X-ray observatories, such as the mission concept study \textit{Lynx} and the selected \textit{Athena}, will require a further shift towards batch/mass production to meet the demand of increased surface area through thin nested mirrors. Batch production will require technologies that can offer adaptability in order to process multiple optical geometries and an ability to be operated in parallel - i.e. to avoid single points/persons of failure. Therefore, the technologies discussed in this chapter all follow these considerations; as a community we are shifting astronomical X-ray mirror production from low volume part counts (units) to `high' volume (batch/mass).      

\section{Evaluating optical surfaces}

This section provides the nomenclature used to describe astronomical X-ray mirror fabrication and the common material characteristics that are desired. The nomenclature is divided into two sections, the first describes the geometry of the mirror, and the second describes how the mirrors are quantified for performance.

\subsection{Terminology: basics}
\begin{figure}
    \centering
    \includegraphics[width=11.5cm]{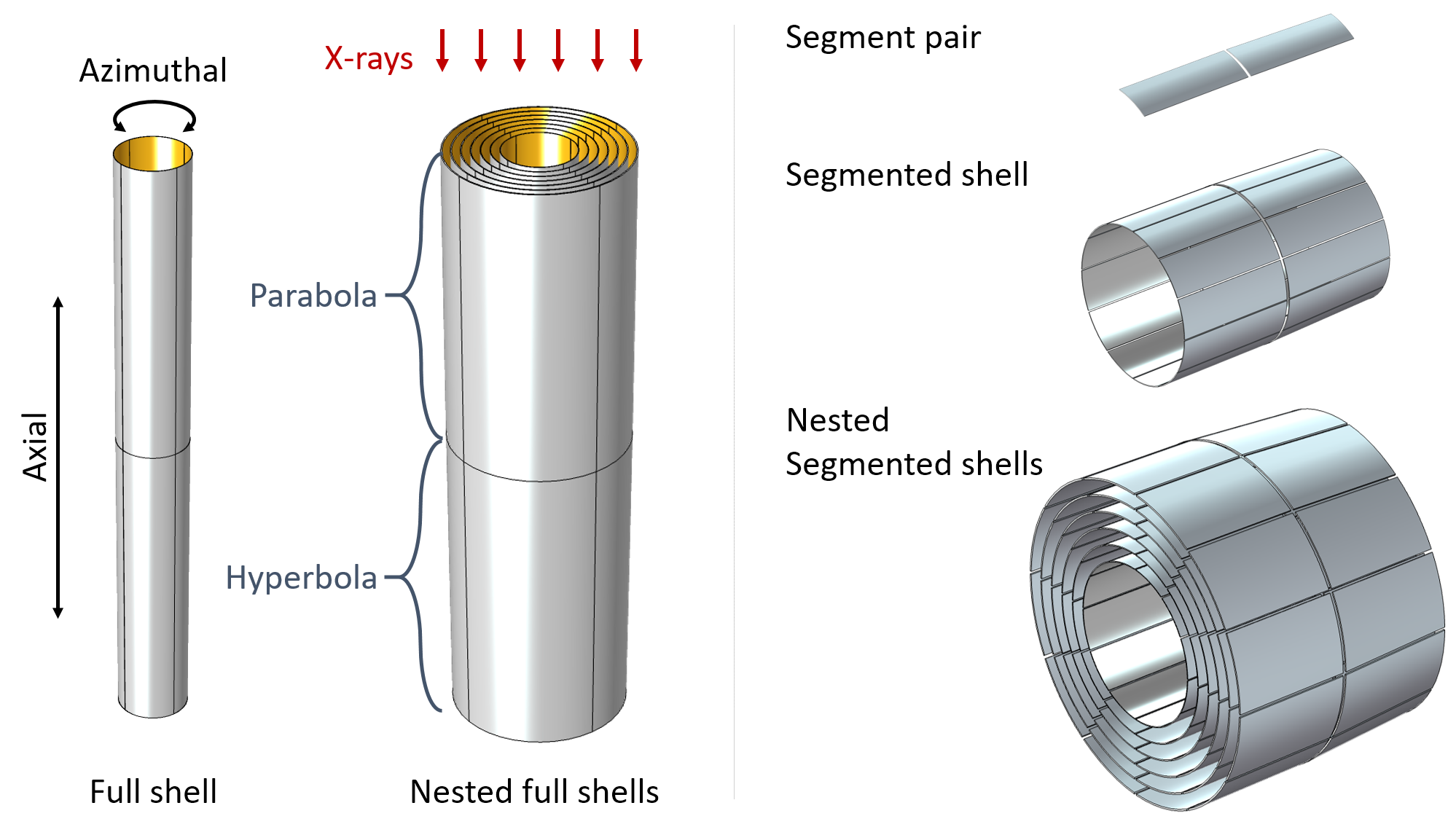}
    \caption{Fundamental X-ray mirror terminology: \textit{left} - full shell (monolithic) mirror shells, individual and nested; and \textit{right} segmented mirror pairs, individual, shell and nested.}
    \label{fig:wolter101}
\end{figure}

Figure~\ref{fig:wolter101} presents the fundamentals in describing X-ray mirrors. X-rays enter the mirror and hit the parabolic mirror of the Wolter I geometry~\cite{WolterH52}, reflect to the hyperbolic mirror and then reflect to a focus, or detector. X-rays travel along the \textit{axial} length of the mirror which has \ang{360;;} rotational symmetry (\textit{azimuthal} direction). Full shell, or monolithic, mirrors have an unbroken azimuthal profile (Figure~\ref{fig:wolter101} \textit{left}). Segmented mirrors are sectors of the full shell, which are aligned in the axial and azimuthal directions to create the full shell (Figure~\ref{fig:wolter101} \textit{right}). Full and segmented shells are \textit{nested} to increase the collecting area.    

\subsection{Terminology: optical surface}
\begin{figure}
    \centering
    \includegraphics[width = 11.5cm]{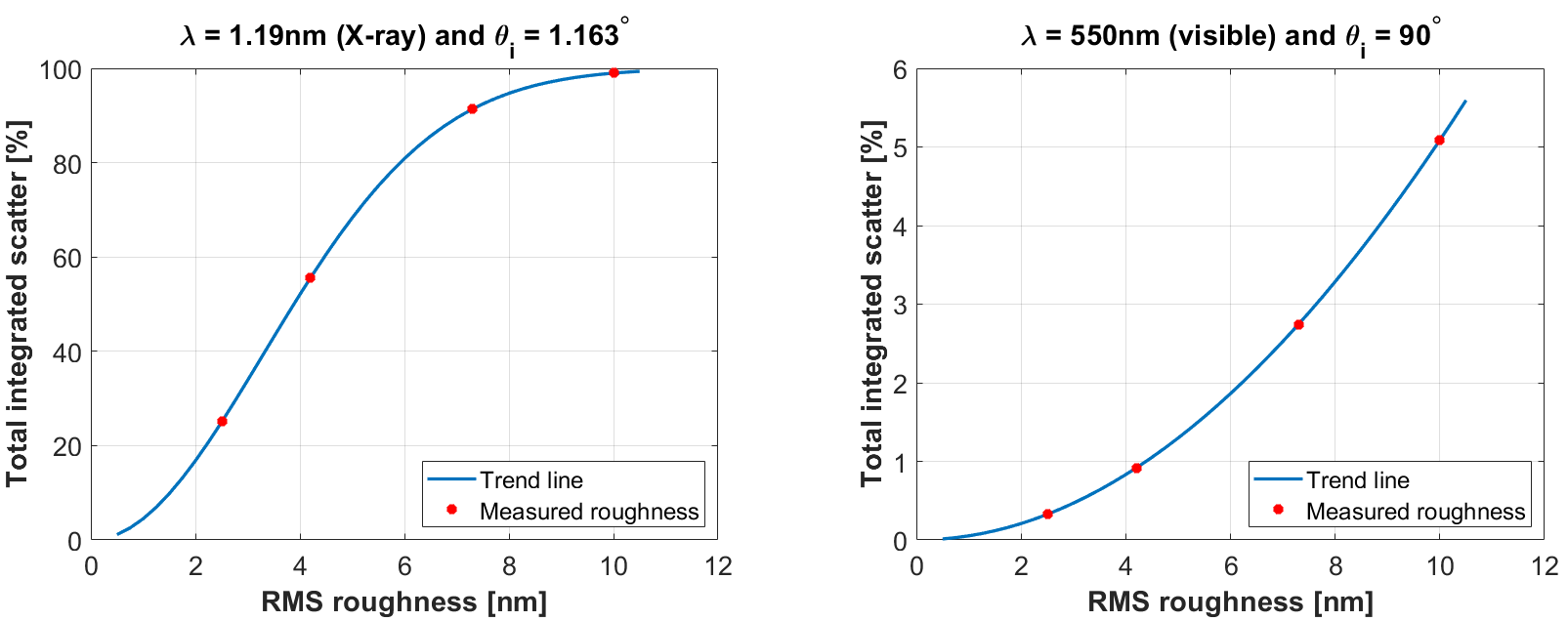}
    \caption{Total integrated scatter as a function of surface roughness at X-ray (\textit{left}) and visible (\textit{right}) wavelengths.}
    \label{fig:TIS}
\end{figure}

The roughness of an optical surface, combined with the angle of incidence and the photon energy, impacts how the incident photons scatter from the surface. In X-ray systems, surface roughness values of less than \SI{1}{\nano\metre} RMS (root mean square) are quoted to minimise the effect of scatter -$~\sim5\%$ scatter at a given wavelength is a typical rule of thumb. The total integrated scatter (TIS) equation, $TIS = 1 - exp(-(4\pi \sin \theta \sigma/ \lambda)^{2})$, can be used to estimate scatter as a function of roughness (RMS, $\sigma$), wavelength ($\lambda$) and angle of incidence ($\theta_{i}$). TIS provides an estimation of the scatter assuming a Gaussian distribution of roughness. Figure~\ref{fig:TIS} demonstrates the effect of surface roughness on the TIS, by comparing X-ray and visible wavelengths. The value of roughness is influenced by the method of manufacture and the material being processed; materials that have a high value of the physical property \textit{hardness} (silicon, quartz etc.) tend to exhibit good roughness once processed.  

\begin{figure}
    \centering
    \includegraphics[width = 11cm]{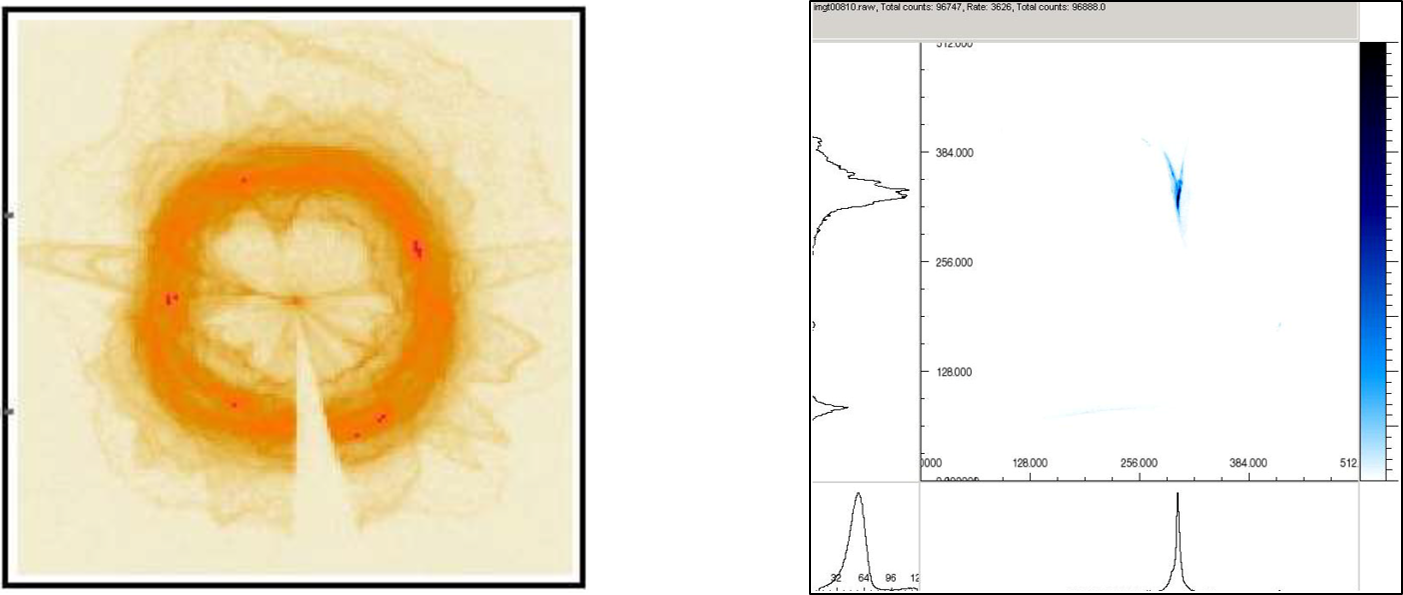}
    \caption{\textit{Left} an out of focus X-ray image highlighting deviations from a perfect ring (image credit: Kilaru, K., et al. 2015~\cite{Kilaru2015}); and \textit{right} an in-focus X-ray image demonstrating a departure from a central focus spot (image credit: Feldman, C., Uni. of Leicester).}
    \label{fig:Ferror}
\end{figure}

The surface form error (figure error) is the departure of the measured optical surface from the optical prescription. The influence of form error within an optical surface is the misdirection of photons from the intended focus and results in blurring of the central focal spot. Figure~\ref{fig:Ferror} provides two examples of form error within measured X-ray data. In Figure~\ref{fig:Ferror} \textit{left}, the image has been taken out of focus to relate the measured optical distortion with the physical location on the mirror. An image of an out-of-focus reflection, in this case, should be a perfect ring; however, as observed, the primary ring is blurred and non-circular, with secondary fainter structure present. In Figure~\ref{fig:Ferror} \textit{right} the X-ray image is in-focus; the departure of the X-ray image from a central point results from form error within the optical surface. Form error is the cumulative effect from manufacturing, material, mounting and environmental errors, and given a common requirement for thin ($<$\SI{1}{\milli\metre}) shells, a challenging parameter to minimise.      

Surface roughness and form error represent the two ends of the spatial frequency (\SI[per-mode = symbol]
{1}{\per\metre}) spectrum; roughness is high spatial frequencies and form error is low spatial frequencies. In between the roughness and the surface form error is the waviness, or mid-spatial frequencies. The range of waviness can either be defined by absolute values, or determined from the geometrical dimensions of the mirror; a broad approximate range is \SIrange[]{0.1}{10}{\milli\metre}. The effect caused by waviness is both misdirection and small angle scatter, an it affects the shape of the PSF. The cause of waviness is typically subtractive methodologies (polishing, grinding \& machining) that imprint a frequency of operation. Therefore, by understanding the method of manufacture, the contribution to waviness, and subsequently PSF, can be estimated.         

\subsection{Materials}

Table~\ref{tab:matprop} presents the commonly used materials for X-ray mirror fabrication and a subset of their physical and mechanical properties. An ideal material for X-ray mirrors has a low density to minimise mass to comply with launch-weight restrictions. The material should be stiff to resist deformations negatively impacting the form error - a high Young's modulus. Furthermore, the material should be thermally stable: it should have a low coefficient of thermal expansion (CTE) to minimise thermal distortions and preferably a high thermal conductivity to allow the mirrors to maintain a thermal equilibrium. In the manufacture of X-ray mirrors for astronomy, the material selection is driven by the process and technology used to create the mirror, which in turn is driven by the telescope and science requirements.

\begin{table}[]
    \centering
    \begin{tabular}{p{4.4cm}p{1cm}p{1cm}p{1.1cm}p{1.6cm}p{1.7cm}}
    
         Property & Nickel & Silicon & Zerodur & Fused silica & Borosilicate  \\
         \hline
         Density [\SI{}{\gram\per\centi\metre\cubed}] & 8.90 & 2.33 & 2.53 & 2.20 & 2.23 \\
         Young's Modulus [\SI{}{\giga\pascal}] & 205 & 160 & 90 & 72 & 64 \\
         CTE [$\times$\SI{e-6}{\per\kelvin}] & 12.8 & 7.5 & 0.1* & 0.55 & 3.25 \\
         Thermal conductivity [\SI{}{\watt\per\metre\per\kelvin}] & 79 & 92 & 1.46 & 1.40 & 1.20 \\
         \hline
         \multicolumn{6}{l}{}\\
         \multicolumn{6}{l}{* - values range between 0.1 - 0.007 $\times$\SI{e-6}{\per\kelvin}}\\
    \end{tabular}
    \caption{Common materials and associated properties used in X-ray mirror fabrication.}
    \label{tab:matprop}
\end{table}

\subsection{Section review}
This section has introduced the manufacturing methodologies; how space-based X-ray telescopes mirrors have been made using these methodologies; the fundamental terminology and metrics used to describe X-ray mirrors; and the physical and mechanical properties of commonly used materials. The remainder of this chapter is divided into into the four manufacturing methodologies used to create X-ray mirrors and the associated technologies and processes. As described, multiple methodologies are employed to create a functional part, the following sections describe, in the context of methodologies, which is the technology and/or process that \textit{most} influences the optical performance. The subtractive methodology is divided into two parts, the first part describes non-material dependant technologies and processes, whereas the second part describes subtractive methods applied to specific material, silicon. The order of the methodologies is outlined below:
\begin{enumerate}
    \item \textbf{Subtractive}: polishing \& ion beam figuring (IBF)
    \item \textbf{Subtractive silicon}: silion pore optics \& monocrystalline X-ray optics
    \item \textbf{Formative}: electroforming \& slumped glass optics
    \item \textbf{Fabricative}: active/adjustable X-ray optics
    \item \textbf{Additive}: additive manufacture
\end{enumerate}

\section{Subtractive}

A subtractive methodology describes the removal of material from a part and it is applicable across all spatial  scales, for example: mill, drill and lathe (\SI{}{\metre} to \SI{}{\milli\metre}), grinding (\SI{}{\milli\metre} to \SI{}{\micro\metre}), polishing (\SI{}{\micro\metre} to \SI{}{\nano\metre}), and IBF (\SI{}{\nano\metre}). This section focusses on how subtractive methods are used to create the mirror surface and how advances in technologies and processes will, it is hoped, push X-ray mirrors to sub-arcsecond resolution with large effective areas in the future. The dominant process described in this section is polishing; it can be considered the foundation of astronomical X-ray mirrors. The discussion on polishing is divided into two parts: first, an introduction to the general process and how it has been applied to date; and second, how technology has advanced to enable optimisation linked with automation. This section concludes with IBF, a process designed to converge to the optical prescription on the nanometer scale. 

\subsection{Polishing - general}

\begin{figure}
    \centering
    \includegraphics[width = 11.5cm]{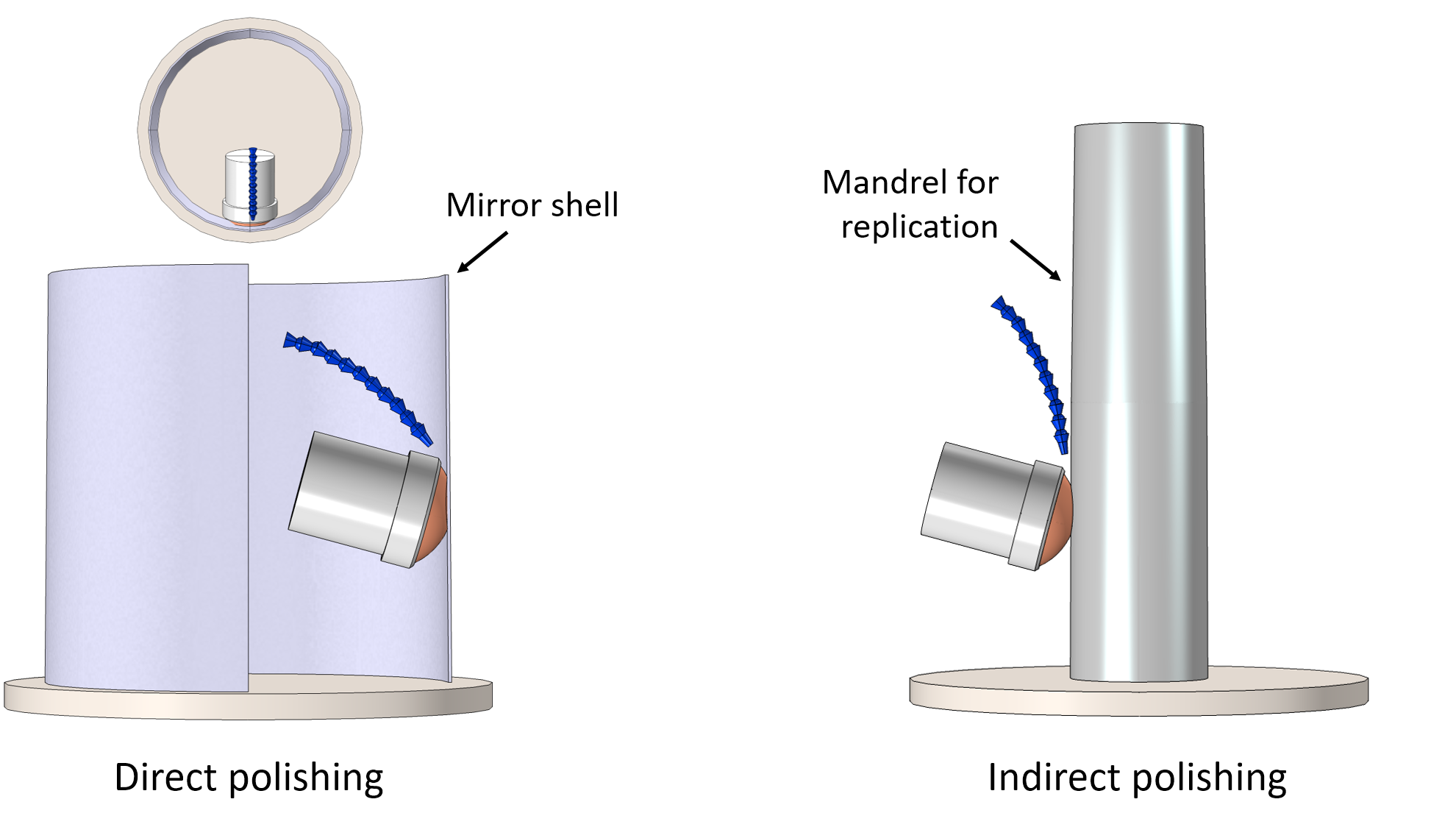}
    \caption{\textit{Left} - an example of direct polishing of full shell optic, the polishing head and slurry need access to the internal surface; \textit{right} - an example indirect polishing, where the external surface of a mandrel is polished for replication.}
    \label{fig:polishing}
\end{figure}

Polishing: the removal of material in a deterministic process using a polishing tool (lap) and an abrasive compound suspended in a liquid (slurry). All the X-ray telescopes described in Table~\ref{tab:xraytele} (and many rocket and balloon X-ray telescope payloads not described) have used polishing somewhere within their production chain, either \textit{directly} or \textit{indirectly} (Figure~\ref{fig:polishing}). Examples of direct polishing include the mirrors for the \textit{Einstein Observatory} (1978), \textit{ROSAT} (1990) and \textit{Chandra} (1999), where the full shell blanks were directly polished to achieve X-ray reflection. Examples of indirect polishing include \textit{XMM Newton} (1999), \textit{NuStar} (2012) and \textit{IXPE} (2021), where a mandrel (a mould) with the inverse shape was polished and the mirrors replicated from the mandrel. In terms of both surface roughness and form error, direct polishing creates a superior X-ray reflecting surface; however, as demonstrated in Figure~\ref{fig:xraymanu}, the trade-off is seen in the requirement for thick, heavy mirror shells that limit the effective area. In contrast, replicated mirrors are unable to mimic the surface quality of the mandrel and therefore there is an increase in the surface roughness and form error. 

Considering full shell mirrors, there are three key challenges for direct polishing: first, the Wolter I optical prescription needs to be polished on the internal surface (Figure~\ref{fig:polishing} \textit{left}), which limits access for polishing hardware; second, to ensure convergence to the desired optical surface, accurate metrology is required and access to the internal optical surface is restricted; and third, a suitable mounting arrangement is required to hold and rotate the shell during polish without imparting distortions onto the optical surface. Direct polishing has demonstrated heritage in delivering sub-arcsecond angular resolution (Figure~\ref{fig:xraymanu}) and therefore it is a `go-to' process where high angular resolution is required. 

\textit{Chandra} achieves its unparalleled X-ray imagery due to the high precision polishing of its four nested mirrors. The thickness of the mirror shells range from \SIrange{16}{24}{\milli\metre} resulting a mirror assembly weighing just less than \SI{1000}{\kilo\gram}; limited by launch-weight restrictions. The increasing shell thickness is proportional to the increasing shell diameter to ensure structurally rigid shells at the larger diameters. Looking to the future, concepts such as \textit{Lynx} (Figure~\ref{fig:xraymanu}) require both precision mirrors and a large number of individual mirror shells to combine both high angular resolution and large effective area. It is not practical to `scale up' \textit{Chandra} and, therefore, a new approach to direct polishing of full shells is required.

\subsection{Polishing - robotic}

Polishing, as an abrasive process, has remained unchanged for centuries; however, advances in polishing as a technology have stemmed from standardisation, synthetic polishing cloths and abrasives, optical metrology and computer numerical control (CNC). The evolution of polishing has progressed from a `by hand' approach, to mechanically driven laps, and now CNC. To date, robotic polishing machines are commercially available that deterministically polish a surface using metrology feedback and knowledge of the polishing tool footprint, or influence function. Similar to conventional polishing techniques, robotic polishing typically requires several iterations to converge on the final optical surface. 

Conventional polishing techniques use a lap with comparable dimensions to the part under polish and the lap will have an approximate inverse geometry to the desired optical profile. Previously for directly polished full shell X-ray mirrors, the lap was translated linearly along the axial length while the shell rotated in the azimuthal direction. Processing an internal surface increases the manufacturing risk and complexity and, furthermore, the unique geometry and optical performance necessitates a bespoke polishing rig. Robotic polishing aims to break the reliance on bespoke hardware, by promoting a single machine capable of polishing a wide range of optical geometries.

\begin{figure}
    \centering
    \includegraphics[width = 11.5cm]{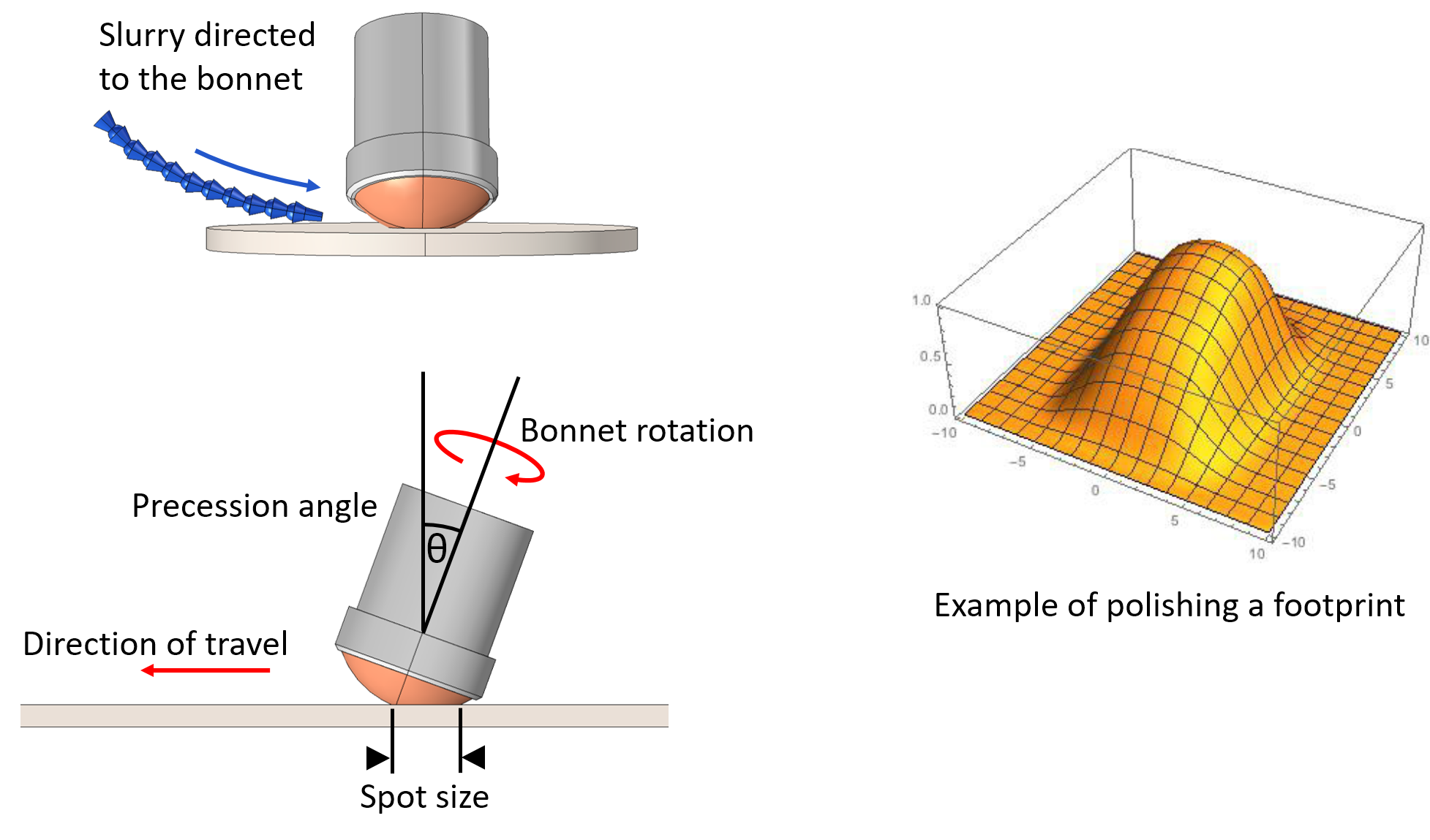}
    \caption{\textit{Left top \& bottom} - a simplification of robotic polishing highlighting some of the parameters to be defined; and \textit{right} an example of a polishing footprint, image credit: Gubarev, M., et al. 2016~\cite{Gubarev16}.}
    \label{fig:RPolishFoot}
\end{figure}

In a simple overview, robotic polishing uses a defined polishing footprint, metrology data of the surface, and the desired optical prescription, to calculate a program (tool-path) that describes the parameters of the polishing operation to iterate from the original to the desired optical surface. The bonnet, an inflated membrane or alternative polishing head, delivers the polishing action to the optical surface. The footprint for a given time period is defined by: the geometric environment of the bonnet (precession angle, offset to the surface), the dynamic environment of bonnet in contact with the optical surface (rotation, translation), and the physical environment (slurry concentration and polishing cloth) - Figure~\ref{fig:RPolishFoot}. The resulting tool-path of the bonnet is unique to the optical surface and therefore accurate indexing of the optical surface to the bonnet is necessary~\cite{Gubarev15}.
 

\begin{figure}
    \centering
    \includegraphics[width = 11.5cm]{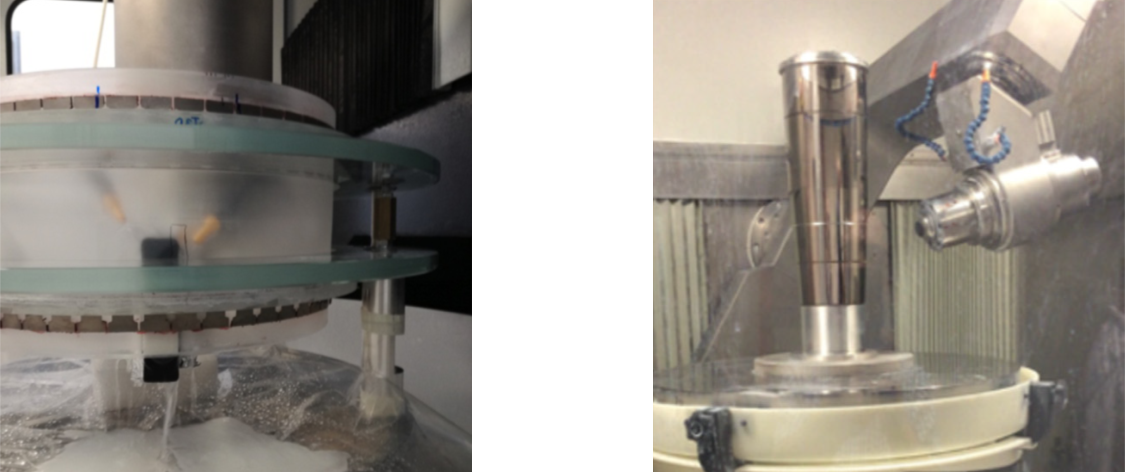}
    \caption{\textit{Left} - the use of robotic polishing applied \textit{directly} to remove mid-spatial frequency errors from a thin fused silica full shell, image credit: Civitani, M., et al. 2019~\cite{Civitani19} (CC BY 4.0); \textit{right} - robotic polishing applied \textit{indirectly} to a mandrel for electroforming, image credit: Kilaru, K., et al. 2019~\cite{Kilaru2019} (CC BY 4.0).}
    \label{fig:robopol}
\end{figure}

Robotic polishing has been used to process X-ray mirrors both \textit{directly} and \textit{indirectly}. Regarding the former, it has been trialled within the production chain of thin ($\sim$\SI{2}{\milli\metre}), fused silica full shell mirrors; a candidate technology for a \textit{Lynx} style mission~\cite{Civitani19}. In this study robotic polishing has been used in two roles, first to improve the roughness of the ground surface to allow for interferometry, and second, to provide a high frequency oscillation smoothing operation of the mid-spatial frequencies (waviness) - Figure~\ref{fig:robopol} \textit{left}. Regarding the latter, it has been implemented to process mandrels for electroforming~\cite{Kilaru2019} - Figure~\ref{fig:robopol} \textit{right}. Mandrels for electroforming, discussed later in this chapter, are solid, metallic, have the inverse form, and are easier to process than thin fused silica full shells. The advantage of robotic polishing in this case is the increased rate of convergence to the desired optical prescription.  

Polishing, in all its forms, is likely to remain the foundation of X-ray mirrors for the foreseeable future; both \textit{directly} and \textit{indirectly}. There is significant heritage in polishing, both in the process and the technology that underpins it. Looking to the immediate future, robotic polishing can be linked with \textit{in-situ} metrology to enable a rapid convergence to the desired optical prescription. Looking beyond, robotic polishing and \textit{in-situ} metrology provide a foundation for an Industry 4.0 approach, where `smart manufacturing', enabled in-part by artificial intelligence, allows for generic machines to deliver a more streamlined and consumer driven product/process - discussed further in the final section of this chapter.



\begin{description}
    \item [\textbf{Advantages}]{Heritage for $<$\ang{;;1} HPD mirrors} 
    \item [\textbf{Disadvantages}]{Slow, metrology feedback, direct polishing of small radii is challenging}
\end{description}


\subsection{Ion beam figuring}
IBF is a subtractive technology to create ultra-precision optical surfaces by removing nanometer scale form errors. The process is applied after final polishing, ensuring that only low amplitude corrections are required, which matches the low removal rate of the process. IBF is a deterministic process that requires prior knowledge of the ion beam footprint, optical surface and desired optical prescription, to calculate the required correction - similar to robotic polishing.

\begin{figure}
    \centering
    \includegraphics[width=11.5cm]{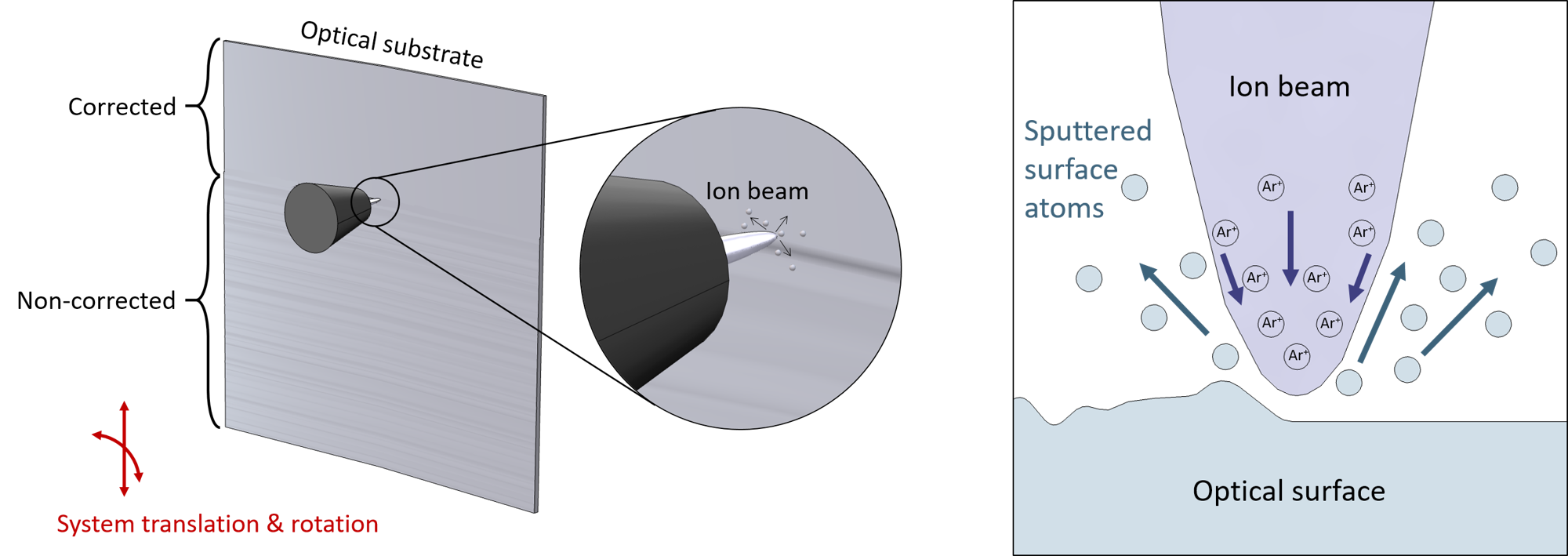}
    \caption{Sketches highlighting the application of IBF to an optical surface.}
    \label{fig:IBF}
\end{figure}

IBF is achieved through the acceleration and focussing of inert gas ions (e.g. Argon, \ch{Ar+}) upon a surface. The impact of the ions on the surface, sputters the surface atoms, leading to removal of material. The shape and amplitude of the ion beam footprint, or removal function, is influenced by the ion mass, energy, incident angle and the surface atoms~\cite{Schaefer2018, Arnold2015}, and typically defined in \SI[per-mode = symbol]{}{\nano\metre\per\second}. The geometry of the footprint at normal incidence is approximately Gaussian and can be readily modelled and verified. To calculate the correction, the amplitude of the ion beam footprints at given locations are calculated from the deconvolution of the optical surface error map and desired optical prescription. The ion beam will dwell for more time in areas requiring the largest correction and less time where minimal correction is required. Figure~\ref{fig:IBF} presents the concept of IBF; rotation and translation of the substrate, or ion beam head, is required to correct the entire surface.

In astronomical X-ray optics, IBF is used to push the surface quality obtained from polishing, to improve the accuracy towards a sub-arcsecond angular resolution. Unless high angular resolution is required, IBF is not a practical step to add in the production chain in terms time, due to its dependence on metrology feedback. IBF has been applied to two variants of segmented X-ray optics: monocrystalline~\cite{Ghigo2018} and slumped glass~\cite{Civitani2018} - methodologies discussed later in the chapter. In the first application, two monocrystalline silicon optics were processed using IBF and demonstrated an improvement in the low spatial frequency ($>$\SI{9}{\milli\metre}) optical form with no degradation to the surface roughness. To implement IBF on a scale compatible with a future flagship mission, efforts are underway to explore how to converge to the optical prescription with a minimum number of iterations; this is critical to ensure that large scale batch/mass production is also cost and time effective


\begin{description}
    \item [\textbf{Advantages}]{Localised corrections to achieve $<$\ang{;;1} HPD mirrors, no degradation of roughness} 
    \item [\textbf{Disadvantages}]{An additional process step, reliant on metrology feedback}
\end{description}

\section{Subtractive: Silicon}

This dedicated section for subtractive silicon technologies is due to the paradigm shift enabled by leveraging the semiconductor industry. Silicon is the dominant material used in semiconductor devices (e.g. computer chips) and therefore high quality silicon wafers and boules, plus the hardware to process them, are readily available at low cost. As presented in Table~\ref{tab:matprop}, silicon has a number of advantages: a density equivalent to glass, a high Young's modulus and good thermal properties. Two different technologies and processes based upon silicon are emerging for use in future X-ray telescopes, silicon pore optics (SPO) and monocrystalline silicon X-ray optics; both are described in this section.

\subsection{Silicon Pore Optics}

SPO represent a transformation in the approach of creating an X-ray mirror. Unlike the other processes and technologies, SPO are modular; they are made from silicon segments that are stacked upon each other and the reflective surface is a series of strips separated by silicon ribs. Once stacked, the silicon ribs create \textit{pores} (channels) along which the X-ray is reflected and, in this way, the individual pores act as X-ray lenslets. Multiple modules, with different radii of curvature, are brought together to construct the primary mirror aperture, a key benefit is the increased rigidity that can be obtained by the optical unit. SPO technology was developed in response to \textit{Athena}, a L-class ESA mission for launch in the 2030s, which aims to achieve an angular resolution of \ang{;;5} and an effective area of \SI{1.4}{\metre\squared} - as highlighted in Figure~\ref{fig:xraymanu}.

\begin{figure}
    \centering
    \includegraphics[width=11.5cm]{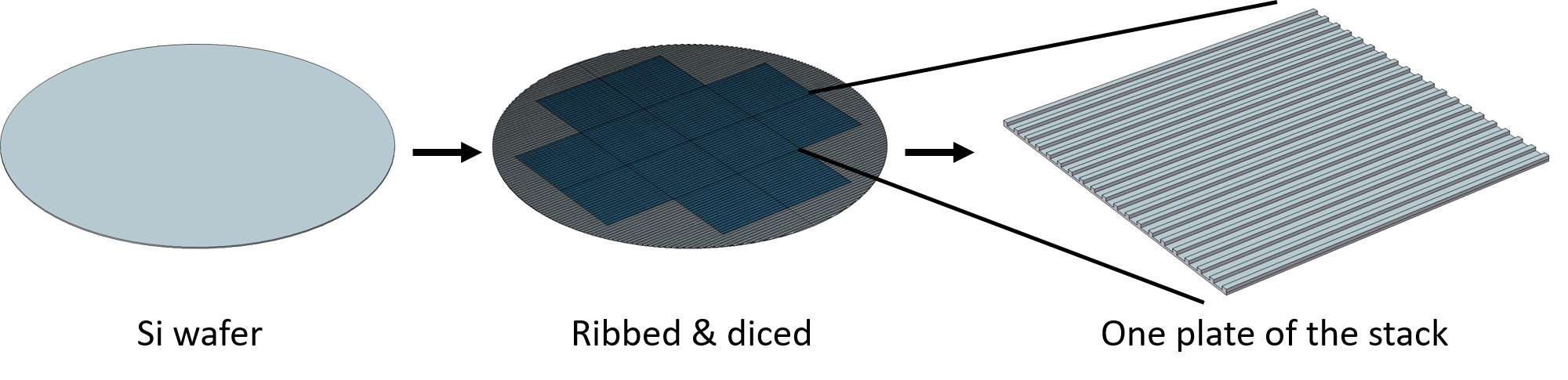}
    \caption{The initial phase of transforming a silicon wafer into a ribbed and diced plate.}
    \label{fig:SPO01}
\end{figure}

Double-sided, superpolished silicon wafers are fundamental to the SPO process. 
These wafers are mass produced, have good surface roughness ($\sim$\SI{1}{\angstrom} RMS), are thin ($<$\SI{1}{\milli\metre}) and exhibit good thickness uniformity. Therefore, double sided, superpolished silicon wafers provide an ideal foundation to build X-ray focussing technology. However, wafers are flat and, therefore, to move from X-ray reflection to X-ray focussing, curvatures in both the azimuthal and axial directions need to be applied. However, stacking wafers without distortion and ensuring alignment is challenging, the SPO approach is to cut the ribs, which form the pores, directly from the silicon wafer, thereby ensuring a single material throughout the SPO module. A summary of the SPO fabrication steps is described below~\cite{Collon2015, Collon2019}: 

The process starts with double-sided, superpolished, \SI{300}{\milli\metre} diameter silicon wafers. The wafers are ribbed (i.e. have channels created on one side) and then diced (i.e. cut) into the required plate dimensions, as shown in Figure~\ref{fig:SPO01}. The optical surface of the plate is the non-ribbed side, which retains its original roughness. 

\begin{figure}
    \centering
    \includegraphics[width = 11.5cm]{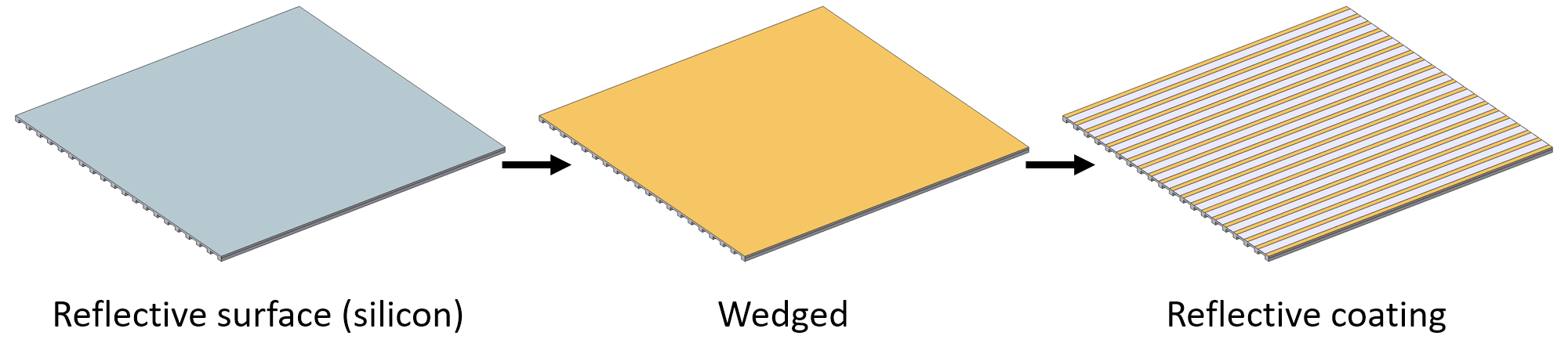}
    \caption{Conversion of the flat reflective surface of the silicon plate into a wedged X-ray reflective surface ready for stacking.}
    \label{fig:SPO02}
\end{figure}

To impart the axial profile, the reflective surface is wedged via a chemical process. The result is that the entrance of the pore will be on a slightly larger radius than the exit, with respect to the central optical axis. Silicon does not demonstrate a good broad band X-ray reflectivity and therefore, an X-ray reflective coating is required.  The reflective coating is deposited in strips at the location of the pores; no coating is deposited immediately above the rib structures to allow for a silicon-silicon bond between plates (Figure~\ref{fig:SPO02}).  

\begin{figure}
    \centering
    \includegraphics[width = 11.5cm]{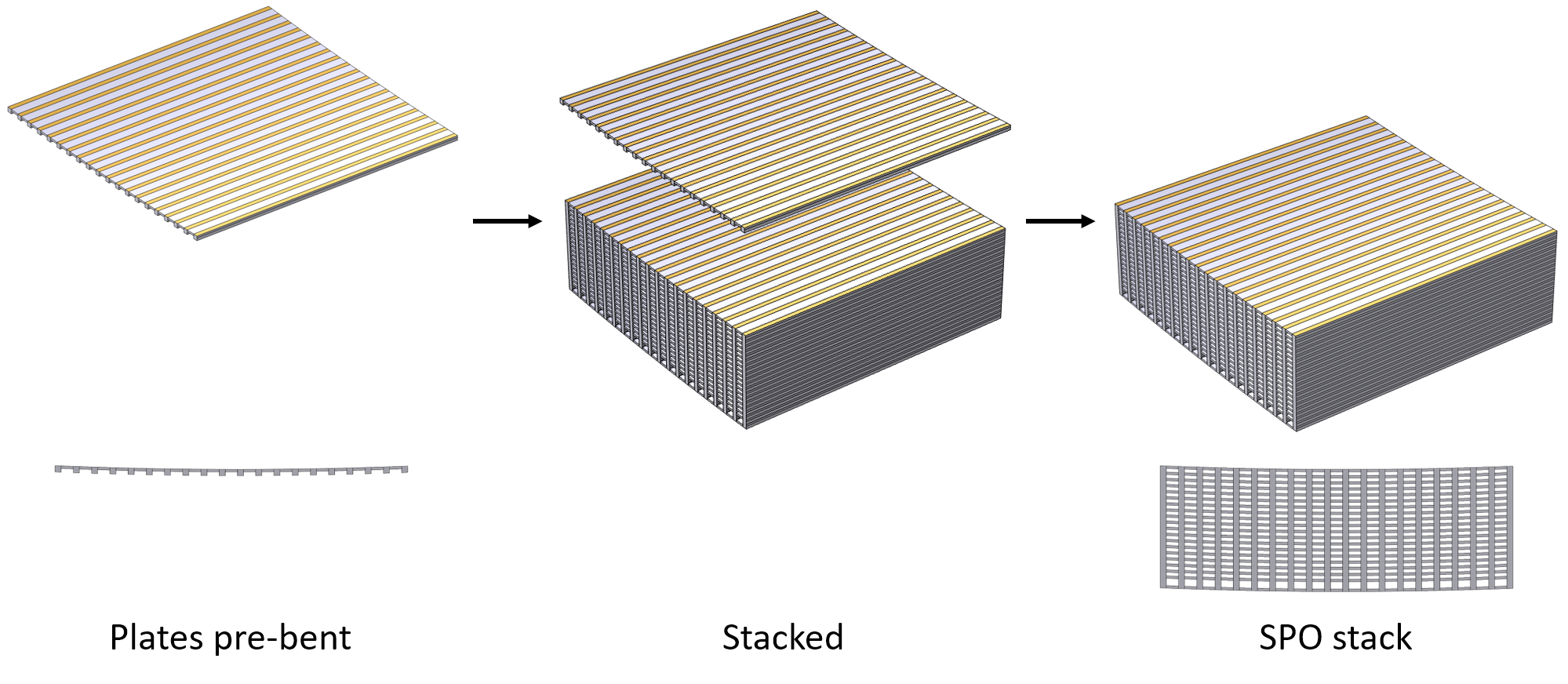}
    \caption{The stacking of the individually curved plates to create the SPO stack.}
    \label{fig:SPO03}
\end{figure}

To introduce the azimuthal radius, the individual plates are `pre-bent' upon a convex die with the approximate radius, during this process the ribs replicate the convex geometry and the optical surface replicates a concave form. To create the stack and impart the desired radius, a high quality concave silicon mandrel provides the foundation. The first plate in the stack is laid with the ribs accessible for bonding, subsequent plates are laid by aligning the mandrel to the die holding the pre-bent plate. The second plate will connect ribs-to-ribs and the third plate will then connect `non-coated wedged strips'-to-ribs - as demonstrated in Figure~\ref{fig:SPO03}.

The key enabling feature of this technology is the connection between plates; as neither adhesives or fixtures are used. The connection is achieved through hydrophilic bonding between the silicon ribs and the wedged non-coated silicon surface~\cite{Collon2015}. The hydroxyl groups that enable the bonding are created through `activating' the bonding surfaces. When the bonding surfaces are brought together, a weak bond is created via Van-der-Waals forces, which is then converted to a stronger covalent bond through an annealing process. Avoiding adhesives and fixtures, minimises the potential for the introduction of distortions in the stacking process.      

\begin{figure}
    \centering
    \includegraphics[width=5cm]{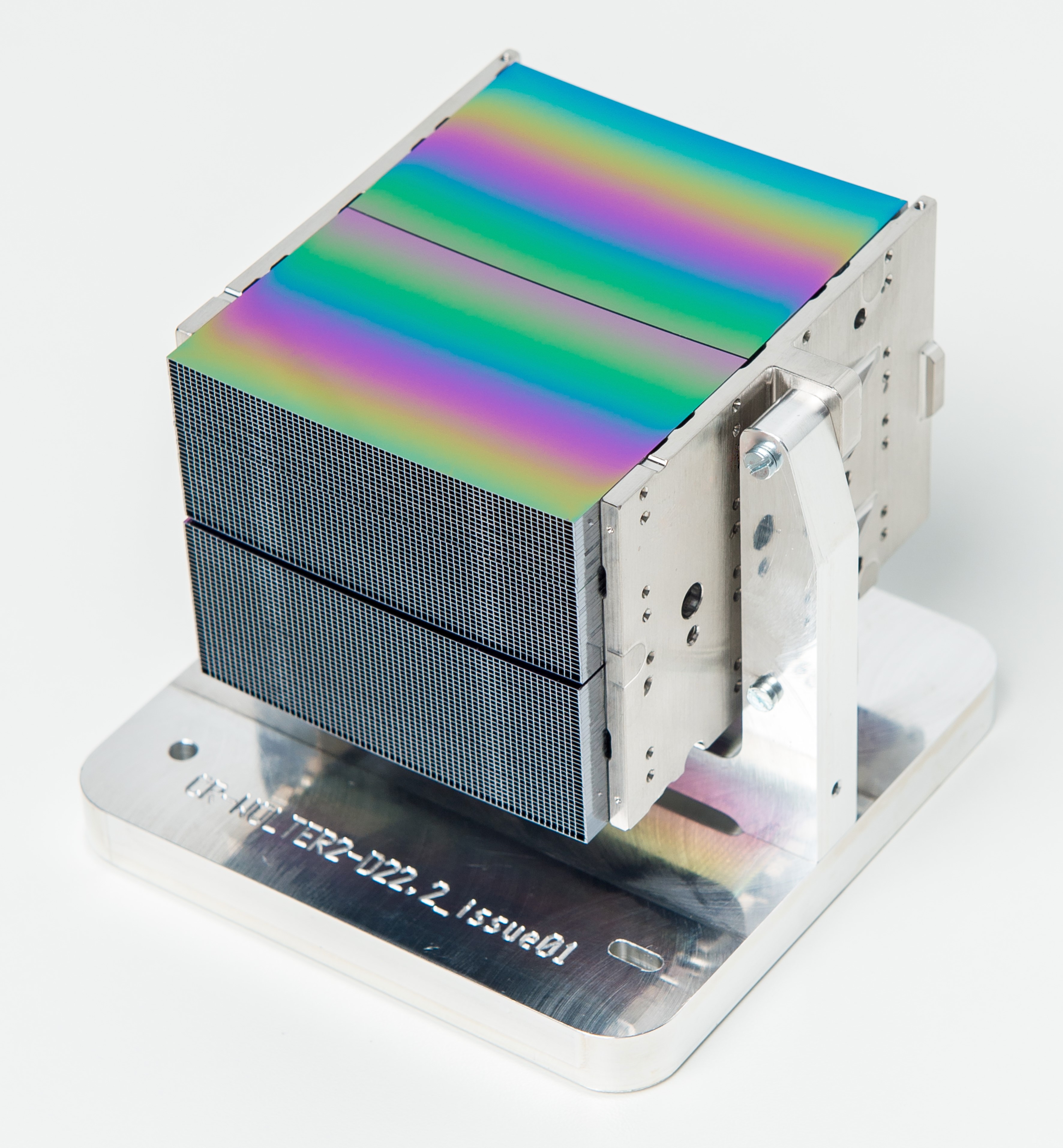}
    \caption{A complete SPO mirror module, image credit: \copyright ~ESA/cosine Research.}
    \label{fig:Cosine}
\end{figure}

A complete stack represents either the parabola or the hyperbola of the optical system, to gain the double reflection of the Wolter I geometry, two stacks are aligned along the optical axis. The two stacks, once aligned, form the mirror module as shown in Figure~\ref{fig:Cosine}.  

SPO not only leverage the semiconductor industry, but due to the high levels of accuracy, cleanliness and mass production required, robots (i.e. industrial automation) are used within the stacking process. This introduction of an automated assembly marks a shift in how astronomical X-ray optics have been made to date - i.e. moving from a person driven assembly line, towards an automated production process. SPO break the \textit{status quo} of conventional lightweight X-ray optics production and recent results indicate that modules are now converging to the \textit{Athena} mirror requirements~\cite{Collon2021}.

\begin{description}
    \item [\textbf{Advantages}]{Leverages semiconductor and automation industries, rigid mirror modules, lightweight.} 
    \item [\textbf{Disadvantages}]{Lacking heritage, stacking alignment.}
\end{description}

\subsection{Monocrystalline silicon meta-shell X-ray optics}

Monocrystalline silicon X-ray optics represent alternative approach to leveraging the semiconductor industry to create low cost, lightweight X-ray optics for astronomy~\cite{Zhang2019}. The objective is to combine sub-arcsecond resolution with large effective area targeting future flagship missions, such as the \textit{Lynx} concept shown in Figure~\ref{fig:xraymanu}. To gain the order of magnitude improvement in angular resolution, in comparison to SPO, blocks of silicon are ground, cut and polished to the parabolic or hyperbolic optical form, as opposed to replicating the form from a mandrel. The advantage of this methodology is that the optical prescription is directly generated and this approach has heritage for sub-arcsecond angular resolution, as demonstrated by \textit{Chandra}. Similar to the SPO, this methodology creates mirror segments and these segments are stacked/mounted to create full revolution `meta-shells'.

\begin{figure}
    \centering
    \includegraphics[width=11.5cm]{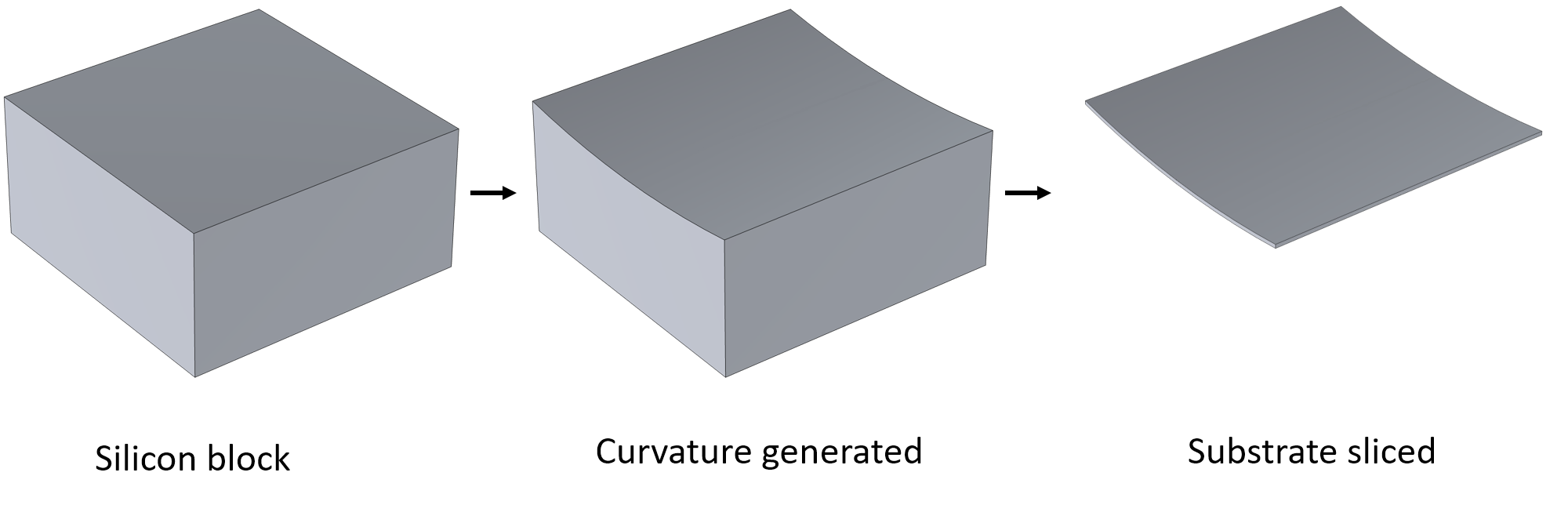}
    \caption{The initial three steps of the silicon segment fabrication, starting with a silicon block, precision machining to generate the conical form and removal of the segment using a slicing saw.}
    \label{fig:Simeta01}
\end{figure}

This approach starts with a block of monocrystalline silicon, which is readily available from the semiconductor industry. First, a conical approximation to the desired optical prescription is machined onto the surface. The conical surface is then removed from the block using a slicing saw, which produces a lightweight conical segment (Figure~\ref{fig:Simeta01}). However, due to the machining and slicing processes, the segment exhibits significant surface damage. 

\begin{figure}
    \centering
    \includegraphics[width=11.5cm]{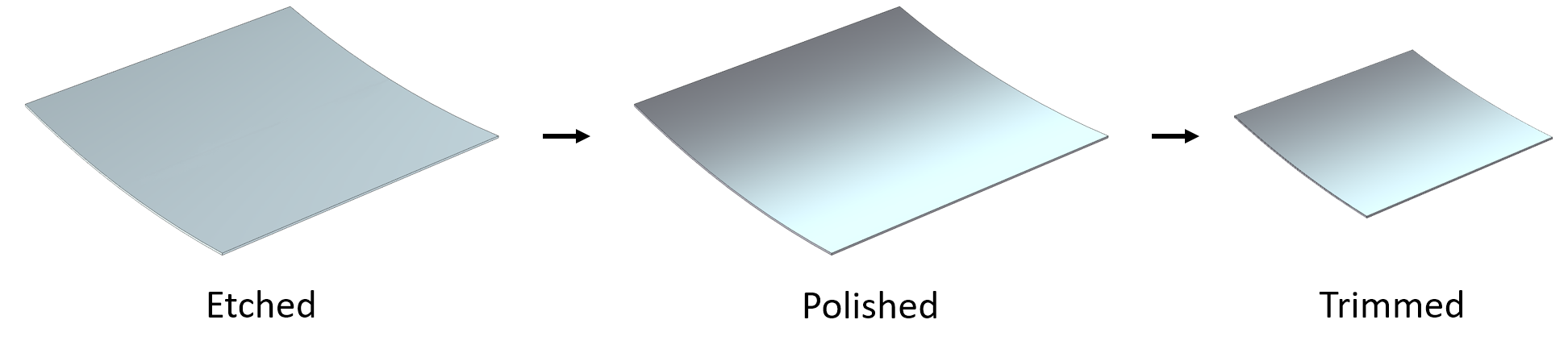}
    \caption{The final steps in silicon segment production: acid etching to remove damage, polishing to achieve surface form and roughness, and trimming to remove edge effects.}
    \label{fig:Simeta02}
\end{figure}

Monocrystalline silicon is an inherently stress free material due to the uniform lattice arrangement of silicon atoms and it is this atomic structure that allows a rough cut segment, to be converted into a superpolished grazing incidence mirror. In the second stage of segment fabrication (Figure~\ref{fig:Simeta02}), the displaced atoms resulting from the surface damage are etched away from the substrate using an acid bath so that only the regular lattice structure remains. The etched substrate then undergoes two rounds of polishing to iterate to the desired optical form (\ang{;;3} to \ang{;;5} HPD) and surface roughness ($<$\SI{0.2}{\nano\metre} RMS~\cite{Zhang2019}). In the final stage of the segment production, the edges of the segment are trimmed resulting in nominal dimensions \SI{100}{\milli\metre} $\times$ \SI{100}{\milli\metre}. In a finishing phase, metrology feedback is used with IBF to converge to the optical prescription, it is suspected that two to three iterations of IBF will push the monocrystalline silicon segments to the \ang{;;0.1} angular resolution.    

\begin{figure}
    \centering
    \includegraphics[width=11.5cm]{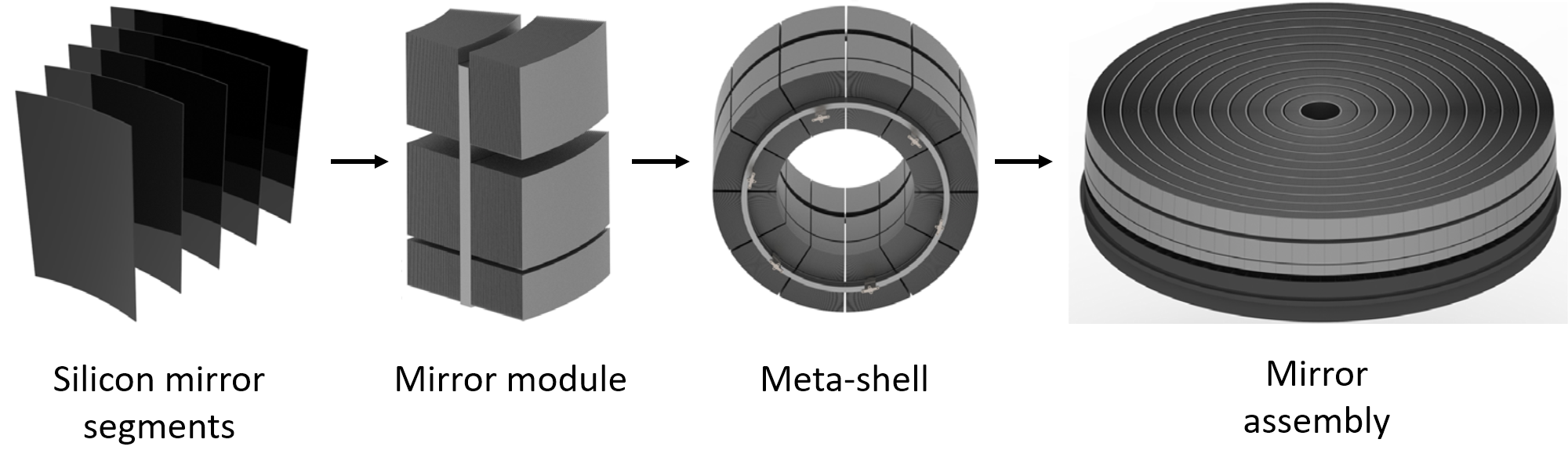}
    \caption{A schematic highlighting how meta-shells can be created from individual segments and brought together within a mirror assembly. Image adapted from Zhang, W., et al. 2019~\cite{Zhang2019} (CC BY 4.0).}
    \label{fig:meta}
\end{figure}

However, the individual segments require mounting and nesting to create the large collecting area. The challenge with a segmented approach is the need to accurately align and nest the individual segments to reproduce a nested full shell geometry. The meta-shell approach divides the telescope aperture into rings (annuli; meta-shells) and each meta-shell has a structural ring that is used to mount the segments upon. Figure~\ref{fig:meta} highlights one design concept for the integration of the mirror segments into a mirror assembly via meta-shells. A key feature of the meta-shell approach is the use of a single material throughout the assembly - i.e. the spacers between segments and the structural ring are also silicon - which ensures good thermal conductivity and minimises thermal distortions. However, unlike SPO, the spacers are bonded to the mirror segments with adhesive, rather than using hydrophilic bonding.      

Monocystalline silicon X-ray optics are targeted towards enabling both sub-arcsecond resolution and a large effective area, which limits the technology transfer between the semiconductor industry and lightweight X-ray optical fabrication. Therefore, there is less opportunity to embed automation due to the bespoke equipment and processes required; however, to enable the production of tens of thousands of mirror segments routes for mass/batch production and an opportunity for automation have been identified~\cite{Biskach2019}. Ultimately, monocrystalline X-ray optics use direct polishing generate the desired optical performance, which is a proven methodology to deliver high angular resolution components (\textit{Chandra}; Table~\ref{tab:xraytele}). However, it is the combination of silicon with a meta-shell approach, which is hoped will enable the game changing union of high angular resolution and collecting area.

\begin{description}
    \item [\textbf{Advantages}]{Combining high angular resolution with a large collecting area.} 
    \item [\textbf{Disadvantages}]{Iterations of polishing/figuring, segment alignment, adhesives.}
\end{description}

\section{Formative}

Formative technologies involve the replication of the optical surface from a mandrel. An advantage is that the mandrel is easier to manufacture and measure than creating the full shell of the Wolter I geometry directly, as the mandrel has the inverse form. A disadvantage is that the replicated mirrors do not mimic the same quality of the mandrel, which results in a increase in surface form error and roughness. However, as observed in Table~\ref{tab:xraytele}, for the past two decades formative technologies have been used exclusively in X-ray space telescopes due to their low mass, predisposition batch- and mass-production, and low cost, but the trade-off is in the achievable angular resolution. Therefore, formative technologies tend to be used for science objectives that prioritise flux - i.e. photon count, effective area - over image resolution. This section will discuss two key technologies and processes for X-ray mirror fabrication, electroforming and slumped glass optics; in addition, differential deposition is discussed as a formative correction solution.

\subsection{Electroforming}

Electroforming has been used in X-ray mirror fabrication for astronomy since the 1980s (\textit{RT-4M}~\cite{MANDELSTAM1982} X-ray telescope onboard the Salyut 7 space station 1982) and with earlier use in X-ray solar applications. In this first application, electroforming was termed galvanoplastics and, in the example of \textit{RT-4M}, the thin (\SI{1}{\milli\metre} or less) electroforms were combined with a thicker epoxy outer-layer to create the X-ray mirror~\cite{Hudec1986}. The current application of electroforming, where thin ($\sim$\SI{1}{\milli\metre}) electroforms are used as the X-ray mirror without an additional outer-layer, has been used since the 1990s (\textit{XMM} 1999) and it is still being used today (\textit{IXPE} 2021). In this section the current application of electroforming, from 1990s onwards, is described.  

The foundation of electroforming is electrolysis, where a reaction occurs upon application of a potential difference (voltage). In metal electroforming (and electroplating), the part to be plated is the cathode, and the metal that is to be plated is the anode. In electroforming the deposited metal (electroform) can be removed freely from the cathode, whereas an electroplate bonds permanently to the cathode. In X-ray mirror fabrication, the electroform is the mirror and the cathode is the mandrel. A solution of positive metal ions and negative ions provides the path for the metal transfer and therefore completes the circuit. 

\begin{figure}
    \centering
    \includegraphics[width = 11.5cm]{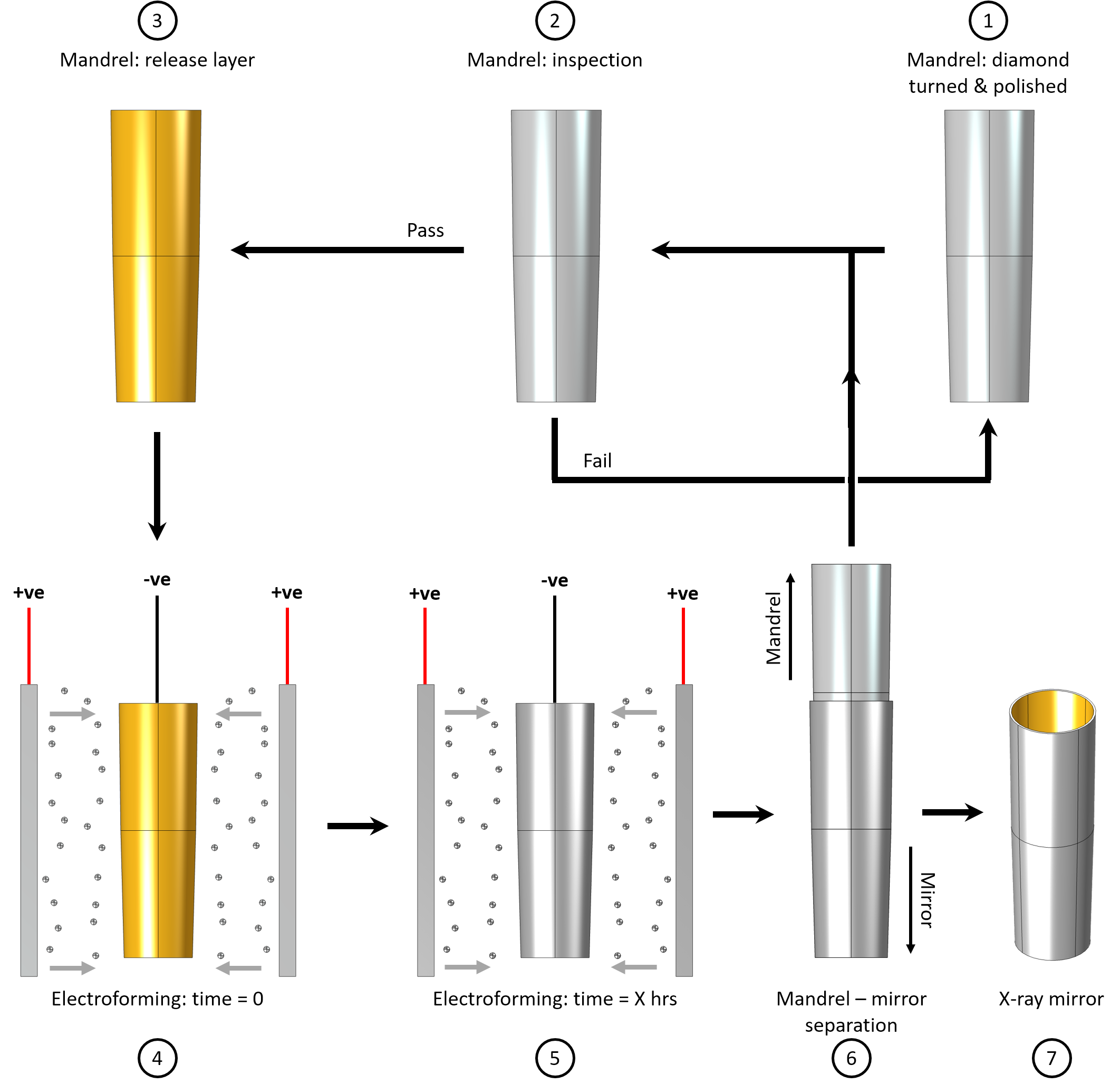}
    \caption{The procedure for the creation of X-ray mirrors by electroforming.}
    \label{fig:electroform_xray}
\end{figure}

Electroforming technology has been used in 6 space-based telescopes to date (Table~\ref{tab:xraytele}) in addition to several rocket and balloon payloads. Nickel, or a nickel alloy, is used as the deposited metal as, under the correct conditions, it exhibits low stress during deposition. Figure~\ref{fig:electroform_xray} provides an overview of how X-ray mirrors are replicated via electroforming. Step 1: the mandrel is diamond turned and polished to the required form and surface roughness. Step 2: the mandrel is inspected to ensure conformance to the required optical prescription. Step 3: a release layer is applied to the mandrel. Step 4: the mandrel is connected within the circuit and situated within the electrolytic solution. Step 5: with the circuit complete and an applied voltage, metal ions are deposited upon the cathode (mandrel). Step 6 \& 7: the mandrel and electroform are removed from the bath and the electroform separated from the mandrel using a thermal shock, at this stage the mandrel can be reused for further replications and the electroform is the X-ray mirror. 

\begin{figure}
    \centering
    \includegraphics[width = 11.5cm]{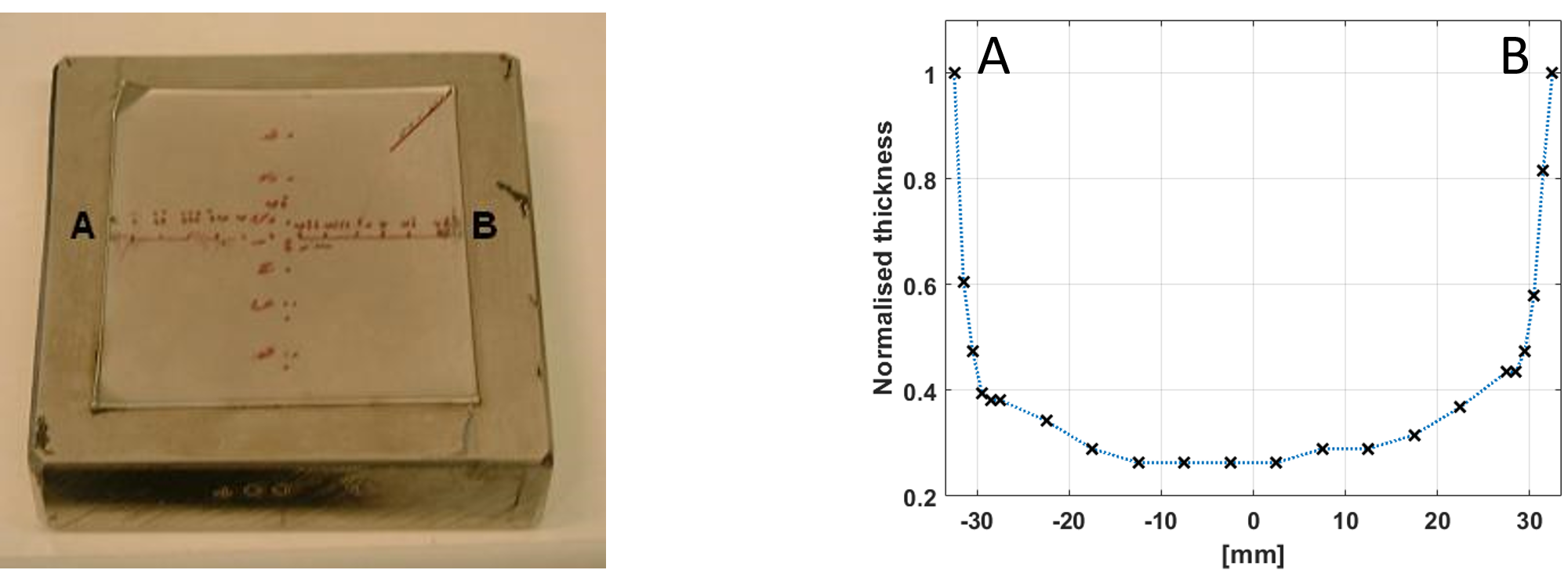}
    \caption{\textit{left} - a photo of a nickel deposit electroformed upon a NiP coated aluminium mandrel; \textit{right} - the thickness distribution of the electroformed nickel.}
    \label{fig:ElectroThickness}
\end{figure}

One challenge in using electroforming is that the thickness of deposition is dependant on the electric field lines - i.e. the current density on the mandrel surface. The electric field lines have a higher density at the edges of the mandrel than in the centre, which results in a non-uniform nickel deposit - as shown in Figure~\ref{fig:ElectroThickness}. To reduce this effect, sacrificial cathodes are used to extend the surface area of the mandrel. Sacrificial cathodes are kept in the plane of the optical surface, but physically separated using an insulating material. Commonly, the sacrificial cathodes are the by-product from the diamond turning and polishing processes where they are used to minimise edge effects. The insulator material must be non reactive within the electrolytic solution and protrude more than the anticipated thickness of the deposit. The insulator can also act to shield the edges of the mandrel, to limit the volume of the electrolyte in the vicinity and thereby the deposition rate.
       
To act as a cathode, the mandrel needs to be metal. Conventionally, it is composed of an aluminium core with the correct nominal form, which is then coated in nickel phosphor (\ch{NiP}) via an electroless process and polished to the required optical prescription. Mirror substrates, across a broad range of wavelengths, are coated in NiP to ensure good surface roughness ($<$\SI{10}{\angstrom} RMS) due to the material hardness and resistance to corrosion. An aluminium core is convenient due to: lower density than \ch{NiP} and therefore lighter to handle; aluminium has a higher coefficient of thermal expansion (CTE) than nickel, which allows the mirror-mandrel separation to occur via thermal shock; and aluminium is less expensive and easier to machine than nickel. 

A second challenge is ensuring that the electroform can be separated from the mandrel - so that it does not become an electroplate. A release layer is used to achieve this and the layer can have one to two functions: for \textit{XMM Newton}, a gold release layer was sputtered onto the mandrel and, upon release, the gold is transferred to the mirror as the X-ray reflective coating; alternatively, for \textit{ART-XC}, the release layer is an oxygen barrier created via passivation of the mandrel prior to electroforming, the X-ray reflective coating is applied in a secondary step.   

Electroforming is an electrochemical process and a successful electroform is dependant upon the chemistry of the electrolytic solution (concentration, pH), physical environment (temperature), as well as the fluid dynamics within the bath (liquid agitation to provide a mechanical flow and to remove bubbles from the cathode). A potential result from non-frequent monitoring is an increase in stress within the deposit, which can lead to an inaccurate replication. The thickness of the deposit can be estimated using Faraday's laws of electrolysis that state that the weight of the deposit is proportional to the applied current. Therefore, by knowing the atomic mass and density of the metal, the applied current, time and surface area, an estimation of thickness can be calculated.   

\begin{figure}
    \centering
    \includegraphics[width = 5.5cm]{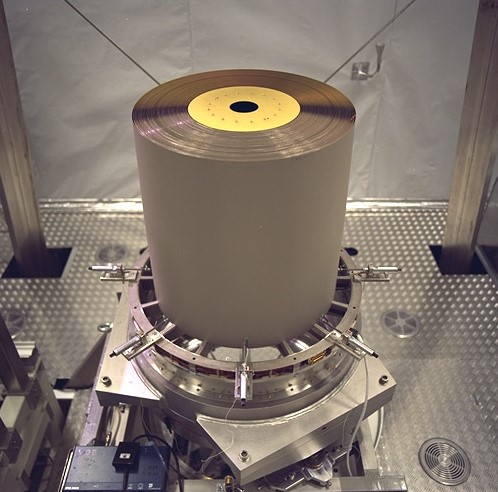}
    \caption{The integration of the nickel elecroformed shells for \textit{XMM Newton}: shell 40 / 58. Image credit: \copyright ~ESA.}
    \label{fig:XMMNewton_assembly}
\end{figure}

Removal of the electroform from the mandrel is achieved through thermal shock. The CTE mismatch is used to separate the two parts, the dominantly aluminium mandrel has a higher CTE than the nickel deposit and, therefore, when submerged within an ice bath the mandrel will contract more than the shell, thereby releasing the shell. The mandrel is then inspected to ensure that it meets the optical requirements, if the mandrel passes inspection it is prepared for further electroforming, if it fails, the mandrel undergoes an iteration of polishing and metrology to meet the optical requirements. The electroformed shell, the X-ray optic, is then either coated with its reflective coating, or inspected prior to integration within the mirror module (Figure~\ref{fig:XMMNewton_assembly}).   

A key limitation of electroforming is the mass of the electroform. The density of nickel is $\sim 4\times$ greater than glass and silicon (Table~\ref{tab:matprop}), which makes electroforming impractical for certain mission requirements. One proposed solution is to electroform a thin layer of nickel ($\sim$\SI{25}{\micro\metre}) and then to overcoat in a thicker alumina layer ($\sim$\SI{200}{\micro\metre})~\cite{Romaine2014}. Alumina has less than half the density of nickel; however, the benefit needs to be balanced with the increase in production steps, which potentially leads to increases in risk, time and cost.  

Electroforming continues to be a `go-to' process for full shell X-ray optics production due to the 10s of arcsecond angular resolution possible and the predisposition towards batch production. Futhermore, electroforming can create shells with a smaller radius of curvature than other methods; however, in contrast, the process is less suitable for large radii of curvature due to the additional mass required to ensure a rigid substrate.  

\begin{description}
    \item [\textbf{Advantages}]{Full shell replication even at small radii, rapid turnaround.} 
    \item [\textbf{Disadvantages}]{Mirror mass, replication errors.}
\end{description}

\subsection{Slumping}

Slumped glass optics are fabricated via a replication process: thin ($\sim$\SI{0.2}{\milli\metre}) borosilicate glass sheets are placed upon mandrels; heated to the glass-transition temperature; become pliant; and, by gravity, take the form of the mandrels beneath. The process is a low cost, rapid turn-around and it was used in the creation of the \textit{NuSTAR} mirror modules. A key advantage is the use of commercially available, off-the-shelf, thin glass sheets with good initial surface roughness. Furthermore, the density of the glass is $\sim$\SI{2.5}{\gram\per\cubic\centi\metre}, which is equivalent to Zerodur and silicon, and therefore offers the potential for lower mass mirror modules than electroforming. 
\begin{figure}
    \centering
    \includegraphics[width=11.5cm]{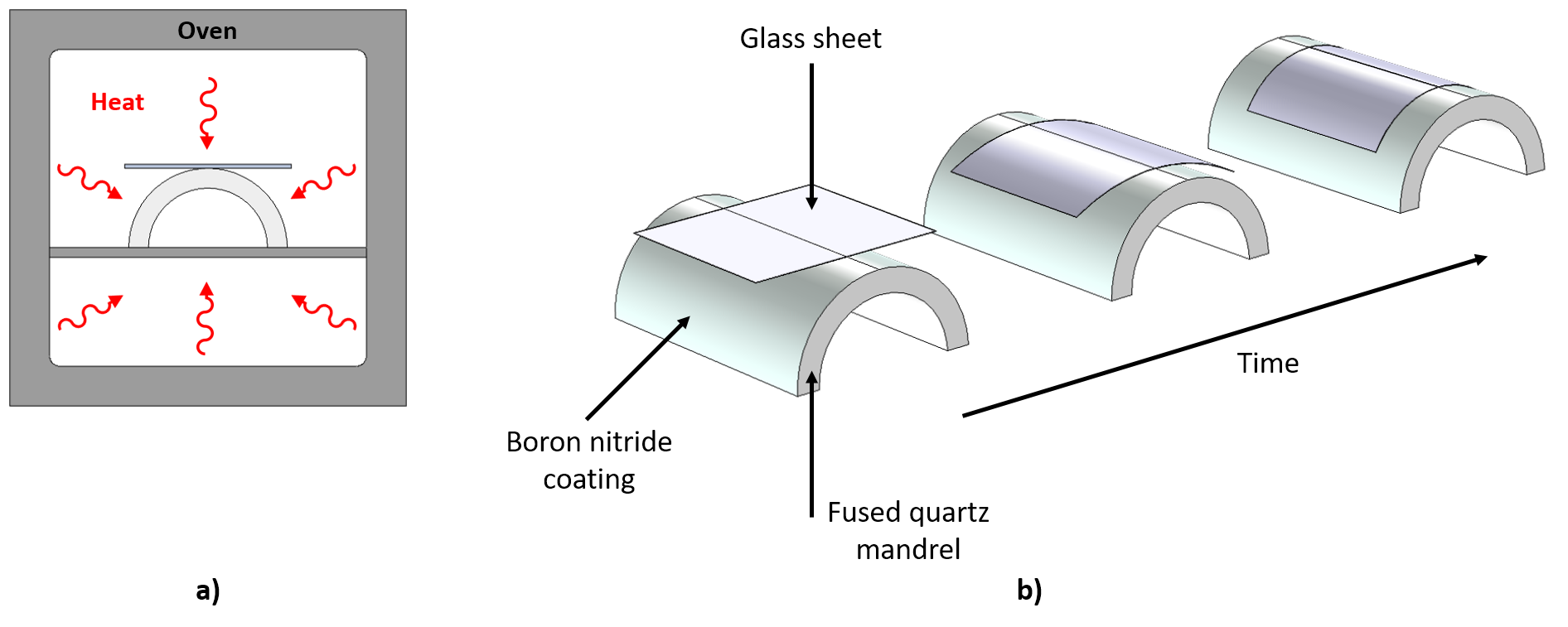}
    \caption{A schematic of the slumping process.}
    \label{fig:slump}
\end{figure}

Figure~\ref{fig:slump} highlights the primary mechanism for creating slumped glass optics and the following description relates to the methodology used in the \textit{NuSTAR} mirror modules~\cite{Craig2011, Zhang2009NuSTAR}. The mandrels start as fused quartz cylinders, which are polished to the optical requirements and then halved; one cylinder yields two mandrels. The mandrels are cylindrical as the conical approximation is generated on the slumped glass during integration within the mirror module. The mandrels are coated in a layer of boron nitride, which ensures that the glass does not bond to the mandrel and can be removed. 

The enabling feature of the process is the off-the-shelf glass sheets exhibiting low surface roughness ($<$\SI{1}{\nano\metre} RMS). Prior to slumping, the glass sheets are cut to the approximate size and cleaned; they are then placed upon the forming mandrels and placed within a oven. The temperature of the oven is increased to $\sim$\SI{600}{\degreeCelsius}, which is approximately the glass-transition temperature ($T_{g}$) where the glass sheet transitions from a hard, brittle state into a pliant state; the $T_{g}$ temperature is less than the melting point of the material. The $T_{g}$ is dependant upon the specific material properties of the glass used. With the oven held at $\sim T_{g}$ the glass deforms under gravity and contacts with the curvature of the mandrel, replicating the cylindrical form. 

\begin{figure}
    \centering
    \includegraphics[width = 11.5cm]{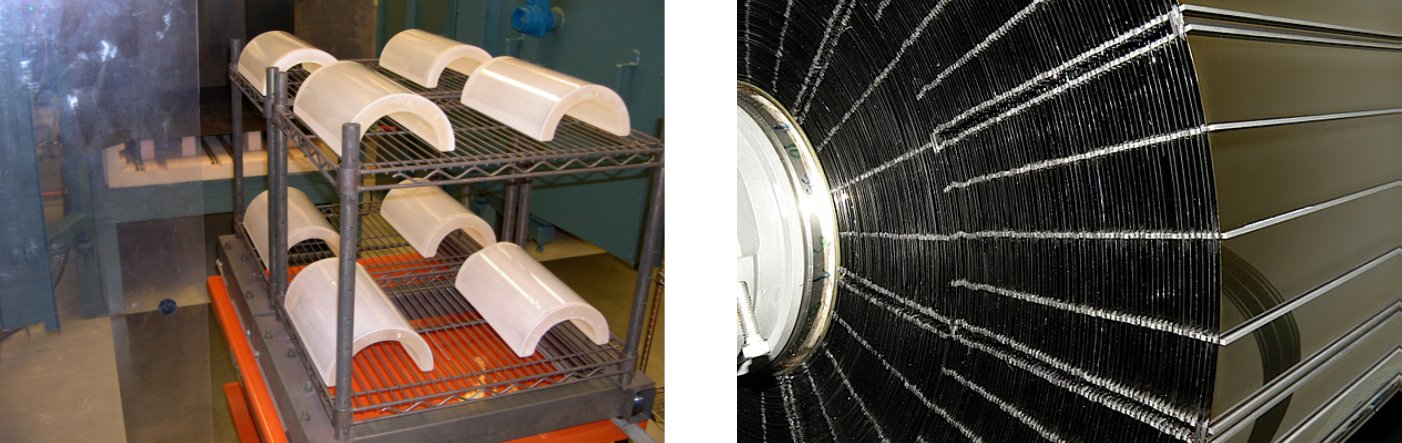}
    \caption{\textit{Left} - slumped glass optics formed upon the fused quartz mandrels, image credit: Zhang, W., et al. 2009~\cite{Zhang2009NuSTAR}; \textit{right} - the integrated slumped glass optics within the NuSTAR mirror module, image credit: Craig, W., et al. 2011~\cite{Craig2011}.}
    \label{fig:GSFC_NuSTAR}
\end{figure}

The oven temperature is decreased slowly to ensure that the slumped glass sheets are annealed to minimise induced stress. The mandrels and cylindrical slumped glass are removed from the oven (Figure~\ref{fig:GSFC_NuSTAR} \textit{left}) and the cylindrical slumped glass substrates trimmed to the required trapezoidal dimensions using a hot wire - the trapezoidal form approximates the conic during integration. From this point, the slumped glass substrates are inspected and progress along the production chain for an X-ray reflective coating and integration (Figure~\ref{fig:GSFC_NuSTAR} \textit{right}). 

For the \textit{NuSTAR} mirror modules, $\sim$\SI{4}{\nano\metre} RMS surface roughness was achieved after slumping. In contrast, the form error from the slumping and integration limited the angular resolution to $\sim$\ang{;;58} HPD. However, due to the success of process and its clear benefits, there have been multiple international efforts to investigate how the process can be improved to yield lightweight high angular resolution optics. For example: slumped glass optics were considered a candidate technology for \textit{Athena}, prior to the selection of SPO~\cite{Salmaso2014}; improvements to the process have been investigated specifically for \textit{adjustable} X-ray optics~\cite{Cotroneo2017} - discussed further in the section on \textit{Fabricative} technologies; and the process has been applied to monocrystalline silicon wafers to take advantage of the mature semiconductor industry post processing steps~\cite{Mika2013}. Finally, slumped glass optics are a candidate substrate for differential deposition, a corrective process where manufacturing defects are corrected through the deposition of a varying thickness thin film, which is discussed in the following section.

\begin{description}
    \item [\textbf{Advantages}]{Inexpensive, rapid turnaround, lightweight.} 
    \item [\textbf{Disadvantages}]{Replication errors, segment alignment to create full shell.}
\end{description}

\subsection{Differential deposition}

Differential deposition can be considered as the inverse of IBF, it is a formative correction process where a variable thickness thin film layer is deposited upon an optical surface to correct for fabrication defects. The goal is to correct for the distortions created via formative X-ray mirror production, electroforming and slumped glass, and thereby improve the angular resolution of the mirror. The process is an additional step within the optics production chain. 

\begin{figure}
    \centering
    \includegraphics[width=8cm]{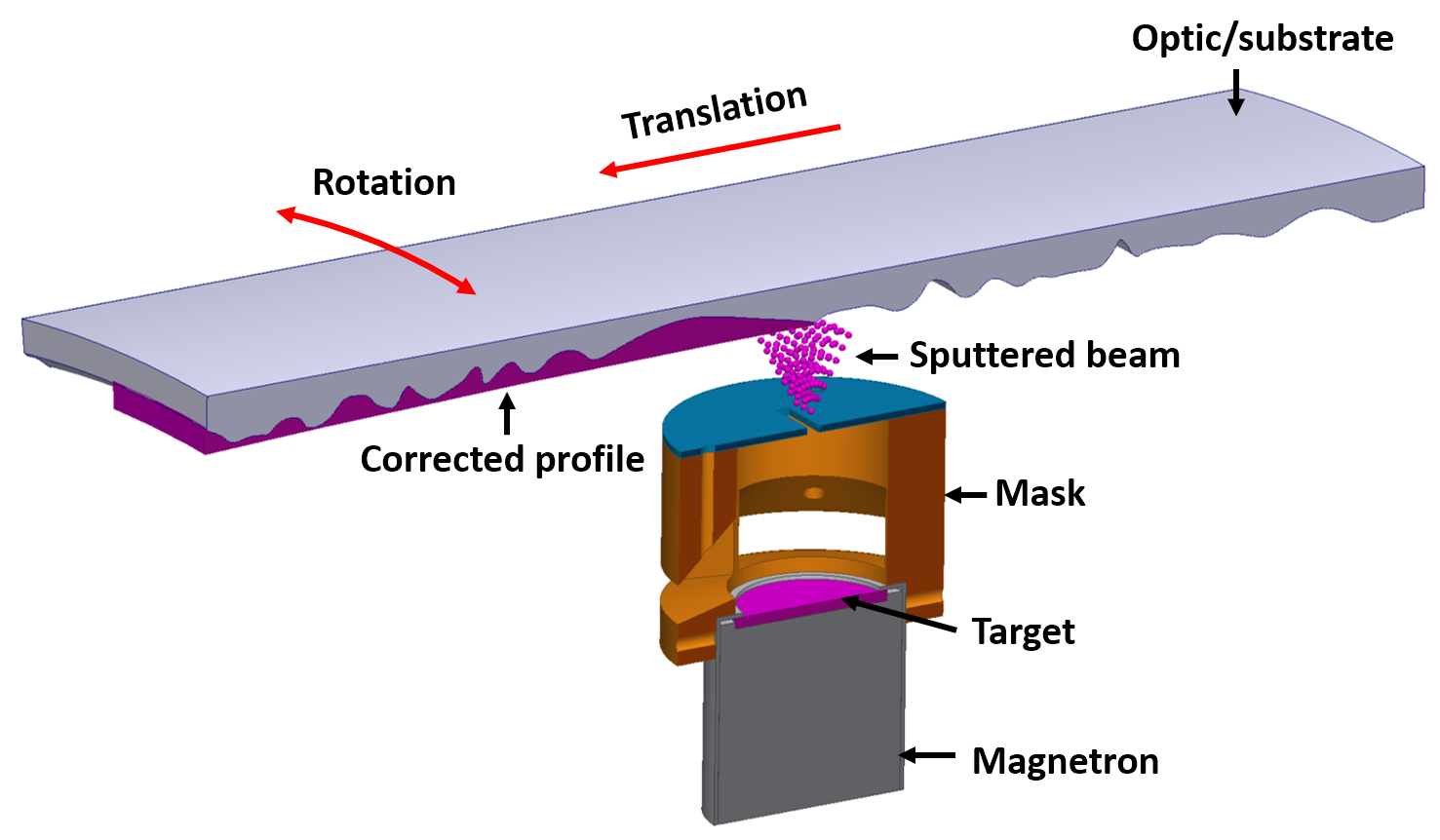}
    \caption{A cross section of the differential deposition process.}
    \label{fig:DD_schematic}
\end{figure}

The concept of differential deposition originates with X-ray mirrors for the synchrotron community~\cite{Alcock2010} and over the past decade, it has been adapted towards thin, replicated astronomical X-ray mirrors. Figure~\ref{fig:DD_schematic} highlights the mechanism for depositing the variable thickness thin film. The substrate translates and rotates by defined displacements relative to a fixed sputtering source and the thickness of the deposition is determined by the time paused at each location - the dwell time. A successful correction of the substrate is dependant upon an accurate deposition profile - i.e. the shape of the sputtered beam at the optical surface - the dimensional metrology used to calculate the correction and the alignment between substrate and sputtering source.

\begin{figure}
    \centering
    \includegraphics[width=11.5cm]{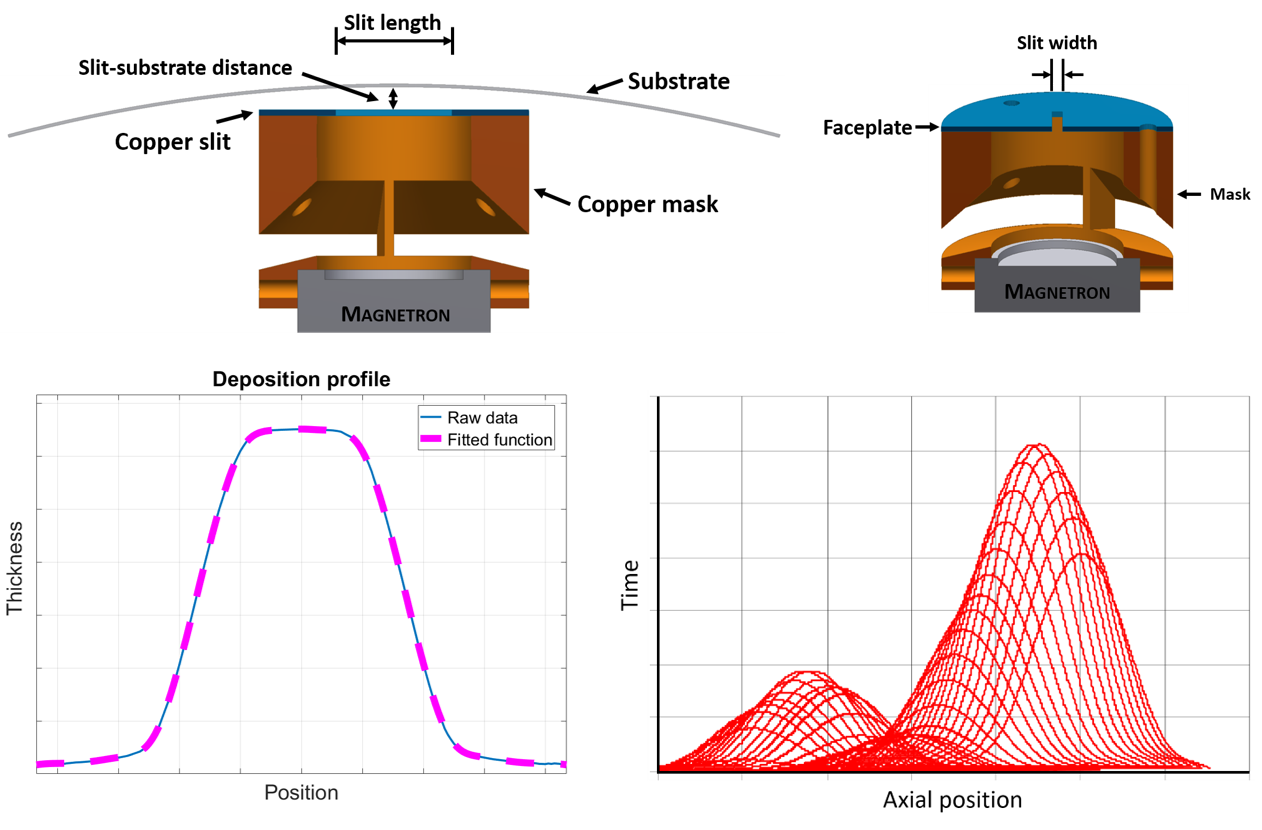}
    \caption{\textit{Upper} - the mask used to define the deposition profile relative to the substrate; \textit{bottom left} - an example of a measured deposition profile (raw data) and the fitted mathematical function; and \textit{bottom right} - an example of the output from the dwell-time optimisation (image credit: Kilaru, K., et al. 2015~\cite{Kilaru2015}).}
    \label{fig:DD_slit_profile}
\end{figure}

A mask with a slit is used to limit the output from the sputtering source (magnetron head) and define the deposition profile (Figure~\ref{fig:DD_schematic}). The shape of the deposition profile is determined by the slit width and the offset of the slit from the optical surface (Figure~\ref{fig:DD_slit_profile} \textit{upper}). To quantify the deposition profile for given slit dimensions and operational parameters (e.g. time, power \& gas pressure), the thickness is sampled across the width of the deposit corresponding to the width of the slit. A mathematical function is used to fit the thickness profile so that the amplitude of the profile at each location can be calculated in the dwell-time optimisation. (Figure~\ref{fig:DD_slit_profile} \textit{lower left}). The dwell-time optimisation is calculated from the metrology of the surface, deposition profile and desired optical profile, as shown in Figure~\ref{fig:DD_slit_profile} \textit{lower right}. The output from the optimisation is a series of locations for the substrate to move to and time the substrate should spend at each location.   

\begin{figure}
    \centering
    \includegraphics[width = 11cm]{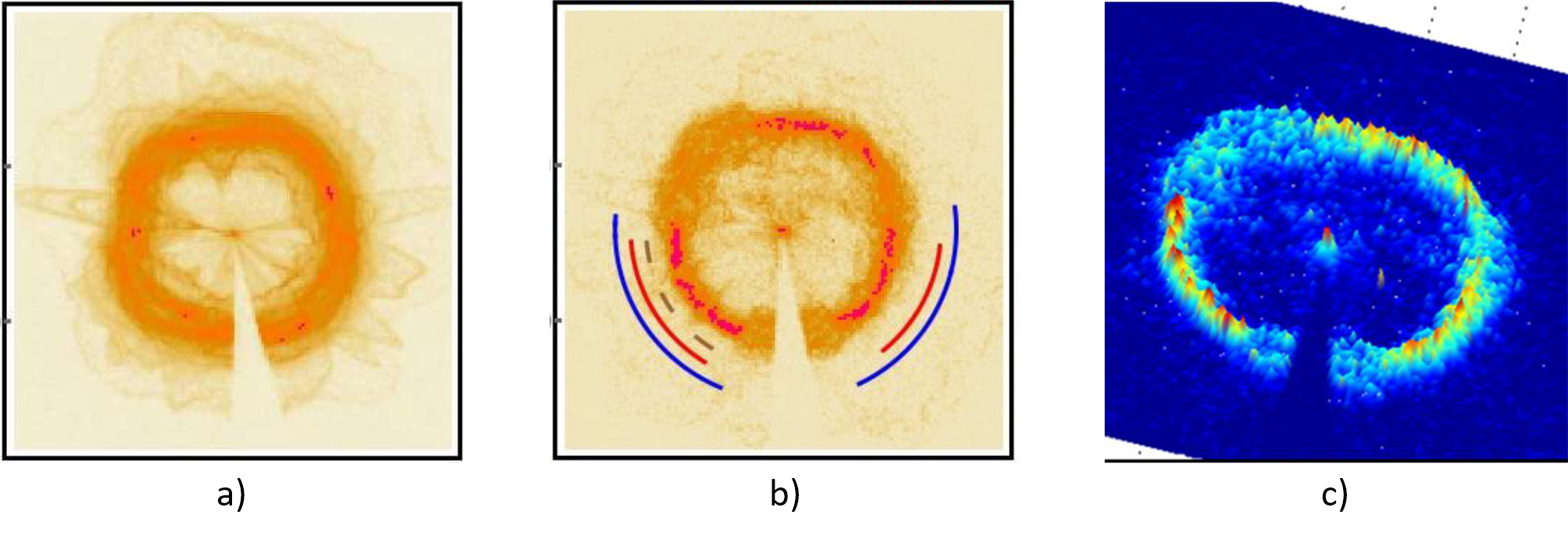}
    \caption{\textit{Left} - intra focus image before correction (the image has been taken out of focus to allow segmental corrections to be observed); \textit{middle} - the intra focus image after correction; and \textit{right} - a 3D representation of the correction in \textit{b)}. Image credit: Kilaru, K., et al. 2015~\cite{Kilaru2015}.}
    \label{fig:DDxray}
\end{figure}

The first application of differential deposition was performed on electroformed full shell nickel alloy mirrors. These mirrors tend to exhibit rotational symmetric errors - i.e. a measured axial profile at \SI{0}{\degree} is similar to profiles measured at \SI{90}{\degree}, \SI{180}{\degree} and \SI{270}{\degree}, and therefore, the full shell can be in rotation at each correction location to achieve a \SI{360}{\degree} correction. Figure~\ref{fig:DDxray} highlights the improvement in X-ray focussing achieved via differential deposition by capturing an intra-focal image created by a corrected mirror. The corrections were applied to $\sim$\SI{60}{\degree} azimuthal segments of the shell to demonstrate the correction relative to the uncorrected regions and to allow different correction strategies to be investigated. The X-ray results demonstrated an improvement in the measured HPD by a factor $>$2 in a single correction~\cite{Kilaru2015}. 

Differential deposition correction has also been applied to replication defects in slumped glass optics. Due to the difference in the manufacturing processes, the slumped glass optics distortions do not have rotational symmetry and therefore require a 3D surface map of each segment. In addition, further challenges are encountered due to the loss of geometric rigidity and a lower Young's modulus, which increases the likelihood of deformations due to stress within the deposited correction~\cite{Atkins2015}. Initial results have demonstrated an improvement in form error along a single axial profile highlighting the potential for correcting segmented mirrors~\cite{Kilaru2017}.


Differential deposition can be applied to improve the figure error of both full shell electroformed and slumped glass X-ray optics and therefore can be considered a candidate technology to merge lightweight and high resolution optical attributes in the future. However, as with IBF, differential deposition is an additional step within the production chain, which inevitably leads to increases in time, cost and risk. Possible routes for reduction of these factors are automated, \textit{in-situ} metrology and accurate deposition profile simulations that are linked with \textit{in-situ} slit width adjustments~\cite{Kilaru2017}.

\begin{description}
    \item [\textbf{Advantages}]{Improving the angular resolution of low cost replicated mirrors.} 
    \item [\textbf{Disadvantages}]{Additional production step, reliant upon metrology feedback.}
\end{description}

\section{Fabricative}

A fabricative approach is one where the image quality of the focussed X-rays is dependant on more than just the mirror surface - i.e. components are added to the mirror substrate to improve performance. Active, or adjustable, X-ray optics are a clear example of a fabricative approach, where actuators are used to deform the existing surface to improve the angular resolution. 

\subsection{Active/adjustable optics}

\begin{figure}
    \centering
    \includegraphics[width=11.5cm]{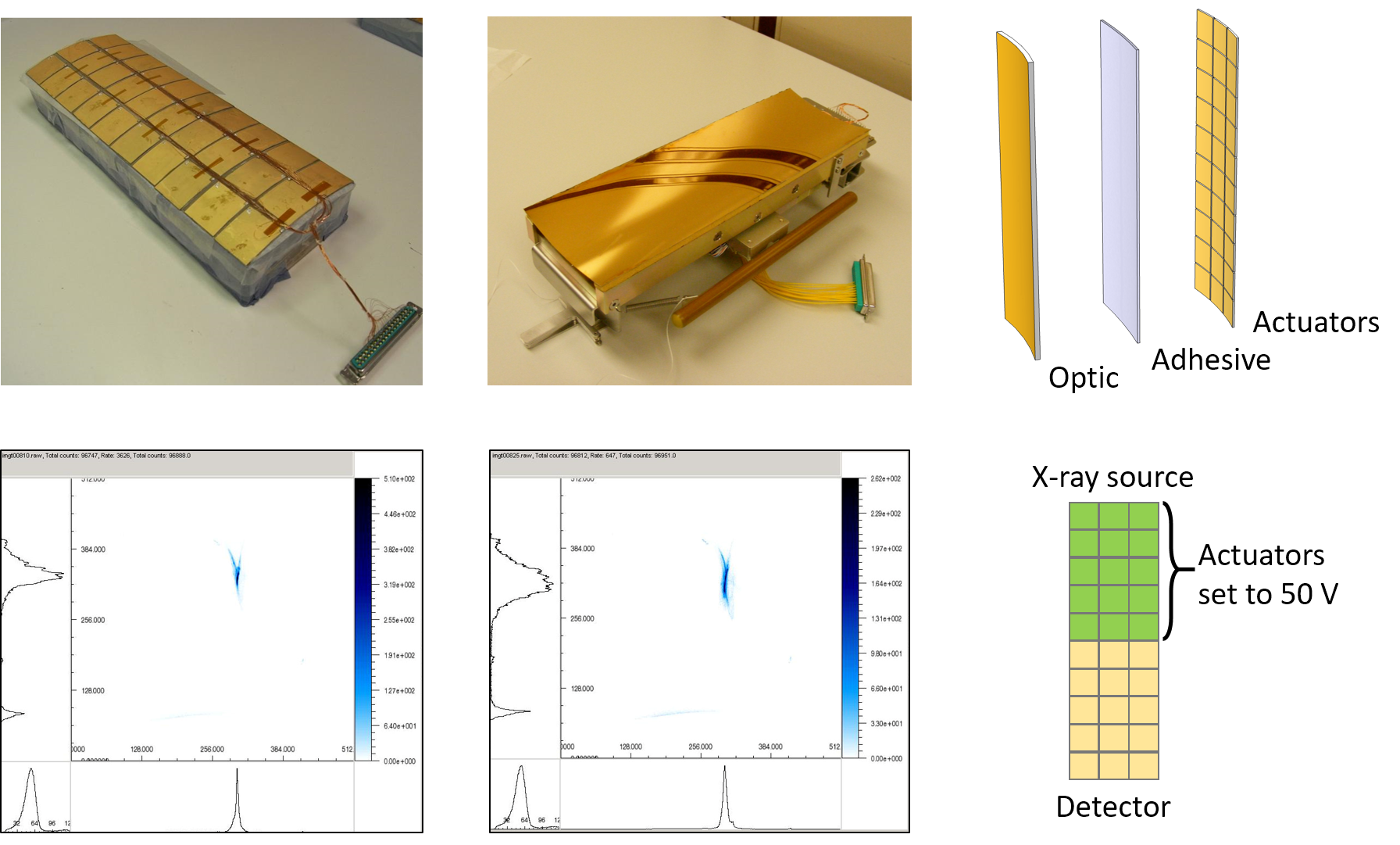}
    \caption{\textit{Upper} - an early active X-ray prototype created by bonding curved piezoelectric actuators upon a nickel ellipsoidal shell. \textit{Lower} - X-ray measurements demonstrating how the actuators alter the focus; \textit{left} before actuation, \textit{middle} after actuation and \textit{right} the actuation pattern.}
    \label{fig:SXO}
\end{figure}

Low- to mid- order spatial frequency errors are targeted with this approach, meaning that the substrates used should exhibit good surface roughness, such as, slumped glass optics. The process is analogous to adaptive optics for ground-based telescopes except that active/adjustable X-ray optics correct for manufacturing defects, as opposed to real time corrections for atmospheric distortions. The process was demonstrated in an early prototype, which used a thin, gold coated, electroformed segment with piezoelectric actuators bonded to the non-reflective surface - Figure~\ref{fig:SXO} \textit{upper}~\cite{Atkins09Prague}. In an X-ray test, the image at the focal plane could be improved using the actuators in terms of HEW and FWHM - Figure~\ref{fig:SXO} \textit{lower}~\cite{Feldman09}. However, the prototype identified several areas for improvement, such as actuator bonding and control, and prototype mounting.   

\begin{figure}
    \centering
    \includegraphics[width=11.5cm]{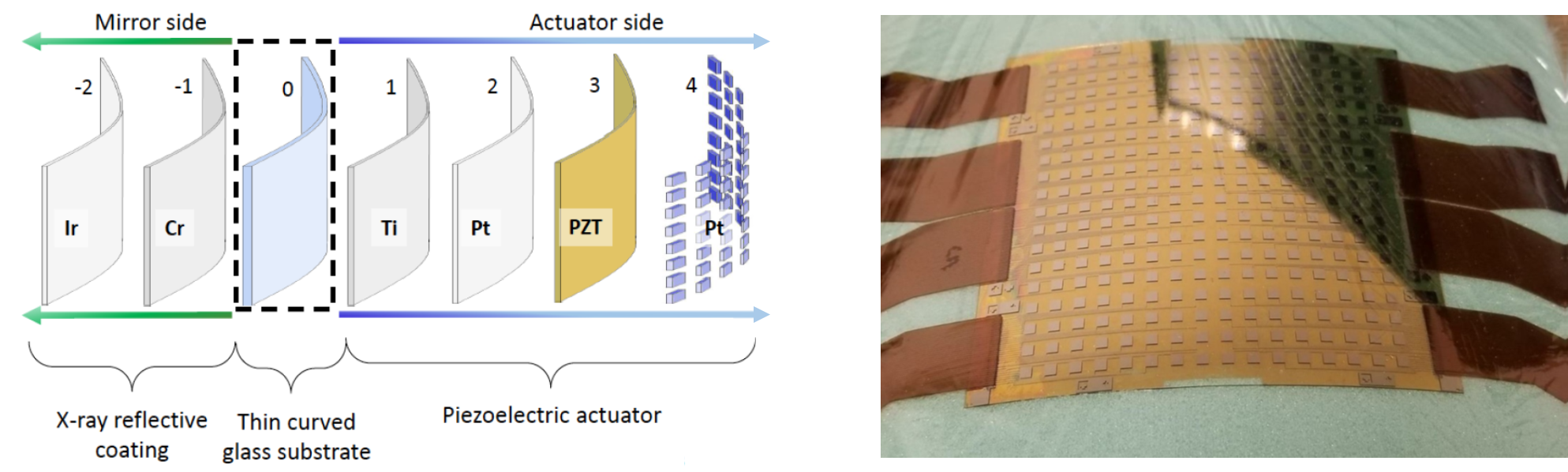}
    \caption{\textit{Left} - the layers that create the adjustable X-ray mirror; \textit{right} - a photo highlighting the prototype mirror. Image credit: Cotroneo, V., et al 2018~\cite{Cotroneo2018}.}
    \label{fig:SAOdiagram}
\end{figure}

Progress over the past decade has targeted the manufacturing challenges of this fabricative approach and leveraged advancements within the Micro-Electronic-Mechanical Systems (MEMS) field. New adjustable prototypes avoid the challenges due to bonding and actuator spacing by depositing the piezoelectric material and electrodes directly upon the optical substrate. In this way, the piezoelectric layer is uniform over the optical surface with a single ground ($\Ground$) and pattered electrodes defining the actuator unit cells. This next generation of adjustable X-ray mirror use slumped glass optics as the optical substrate and Figure~\ref{fig:SAOdiagram} highlights the new multi-layer approach in fabrication.    

However, the direct deposition of piezoelectric material upon the glass substrate requires an extra crystallisation step to convert the sputtered ceramic into a material with piezoelectric properties~\cite{Walker2018}. The temperature of the crystallisation step (\SI{650}{\degreeCelsius}) necessitates an alternative glass substrate with a higher $T_{g}$, in comparison to the \textit{NuSTAR} optics, to ensure crystallisation does not adversely affect the optical form. In addition, to improve the slumped glass process for adjustable X-ray optics, a silicon mandrel is used to form the glass to minimise errors due to CTE mismatch with the new glass. The challenge of stress induced upon the optical surface by the high temperature crystallisation of the piezoelectric material is counteracted by the inverse stress created in the \ch{Cr}/\ch{Ir} layer. Metrology of the optical surface highlights the capability of the new approach to correct for distortions in form error~\cite{Cotroneo2018}.      

The process described above highlights one effort to advance this fabricative methodology. In an alternative approach, electrode patterns have been deposited upon the slumped glass convex surface by photolithography and commercial piezoelectric actuators bonded to the electrodes~\cite{Spiga2016}. In this process, stacking of the active slumped glass substrates was also considered. In a similar approach, commercial piezoelectric actuators were bonded upon the reverse of flat, rectangular, silicon wafers and used, in a Kirkpatrick Baez configuration, to generate the required curvature. X-ray measurements of this system demonstrated an improvement in the FWHM angular resolution using active control~\cite{Hudec2013}. However, piezoelectric actuators are not the only technology capable of imparting deformations upon an optical surface. A different method is by applying a magnetostrictive stress using a magnetic smart material (MSM) within magnetic field. The deformation is then preserved using a magnetically hard material~\cite{Ulmer2017}. In this method, the MSM and magnetically hard material are deposited onto the convex surface.     

The benefit of active/adjustable X-ray optics is the ability to use a low cost replication method to create the optical substrates and to use actuators to deform the surface to the required optical prescription. The technologies and processes that enable active/adjustable X-ray optics, borosilicate glass and smart materials, are commercially available with existing research heritage. Furthermore, due to the corrective ability of the process, there is a lower demand on metrology if it can be assumed that the active/adjustable mirror can correct itself. However, ultimately, the fabricative methodology does add more production steps and potentially necessitates the need for power from a satellite to perform the correction(s) - both adding risk.    

\begin{description}
    \item [\textbf{Advantages}]{Correction of lightweight, low cost substrates towards $<$\ang{;;1} HPD.} 
    \item [\textbf{Disadvantages}]{Increased production and operational steps, increased risk.}
\end{description}

\section{Additive}

Additive is the final manufacturing methodology considered in this chapter. It is most commonly used to describe additive manufacture (AM; 3D printing), but not exclusively, for example, the silcon-silcon bond between the plates in the stacking phase of SPO, demonstrates an additive methodology. Essentially, an additive methodology is one where the objective is to build in layers of the same material. SPO have been previously discussed, therefore this section will speculatively consider AM as a candidate technology for X-ray optics.   

\subsection{Additive manufacture}
AM is a disruptive technology that builds an object, layer-upon-layer, from a digital design file. There are nominally seven different technologies that can offer additive manufacture, such as powder bed fusion, stereolithography and fused deposition modelling. AM is an active area of research for lightweight mirror technology~\cite{Atkins2019a}, with key benefits such as: the creation of lightweight structures that are lighter and stiffer than conventional counterparts; and the ability to integrate additional functionality (i.e. mounting structures, thermal channels) with no additional production steps~\cite{Atkins2019b}. However, there are many challenges to overcome in AM mirror development, including: design optimisation strategies, material property optimisation, space heritage and a lack of suitably trained engineers.

Creating normal incidence lightweight mirrors for ultraviolet, optical and infrared wavelengths using AM has clear benefits and the required optical performance can already be achieved for the longer wavelengths. However, the thin, lightweight, grazing incidence mirrors required for X-ray telescopes are not immediately matched to AM. For example, the surface roughness of AM is microns on a pre-machined surface and therefore subtractive processes are still required to create the reflective surface~\cite{Atkins2017}. Furthermore, thin substrates, a requirement of X-ray mirrors, are prone to stress during printing, which leads to distortions and cracking.

\begin{figure}
    \centering
    \includegraphics[width=11.5cm]{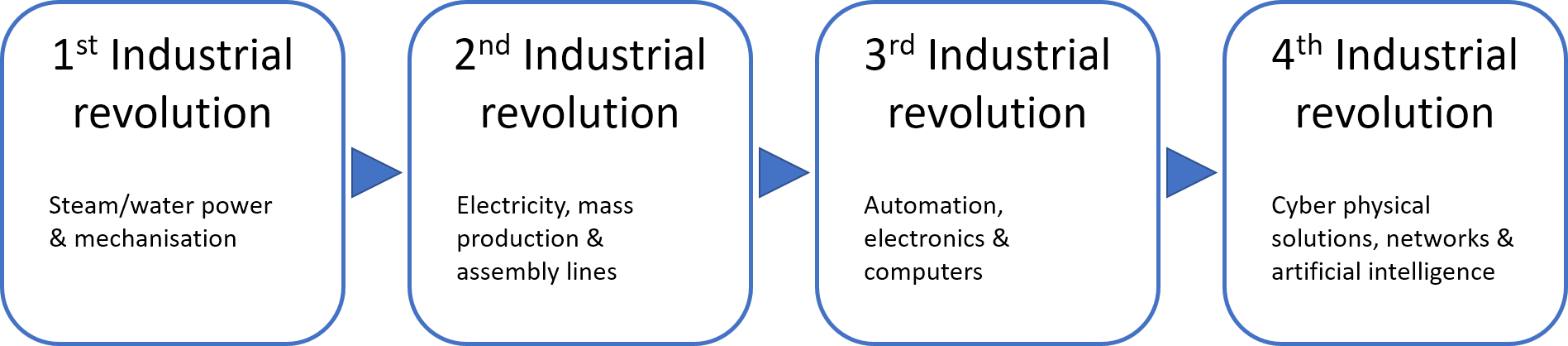}
    \caption{The four stages in the industrial revolution.}
    \label{fig:IR4}
\end{figure}

However, AM is a technology and process to be used, it has clear benefits in a production chain and, with high levels of industry investment, improvements in print resolution and material options will follow. Furthermore, AM is a central technology of the fourth industrial revolution, Industry 4.0~\cite{Vaidya2018}, which envisages design and manufacturing processes linked with smart technology for more streamlined and consumer defined products (Figure~\ref{fig:IR4}).    

Despite the immediate limitations of using AM for astronomical X-ray optics, there are other areas of the X-ray optical production chain where AM has clear benefits, such as within mounting structures: offering the opportunity for part consolidation (i.e. combining multiple individual parts into a single component) and mass reduction. In addition, the range of materials is expanding, technologies to print fused silica~\cite{Kotz2017} could potentially offer a route to print the near-net-shape thin glass shells for polishing, reducing the material to be removed in the grinding and polishing phases. Finally, in an increasingly environmentally conscious society, where there is an expectation of embedding sustainability within production chains and product life cycles, AM offers the community the opportunity to create structures with reduced waste, more function optimised and at a reduced weight.  

\begin{description}
    \item [\textbf{Advantages}]{Part consolidation, mass reduction, design optimisation} 
    \item [\textbf{Disadvantages}]{Lack of heritage, no standardisation, variable materials properties}
\end{description}

\section{Summary}

The goal of this chapter was to provide the reader with an overview of the different technologies and the processes used to create mirrors for X-ray telescopes. The objective of the chapter was to present this field in the framework of the manufacturing methodologies (subtractive, formative, fabricative \& additive) and how these methodologies influence the mirror attributes (angular resolution and effective area). The objective has been supported by fundamental basics in manufacturing, previous X-ray telescopes, and X-ray mirror and optical performance terminologies. The key takeaways from this chapter on \textit{Technologies for advanced X-ray mirror fabrication} are: 1) X-ray mirror fabrication for astronomy is challenging and historically there has always been a balance between angular resolution and effective area; 2) there is more than one method to create an X-ray mirror and, as a result, it is a diverse field of processes and technologies; finally, 3) Industry 4.0 has the potential to disrupt the \textit{status quo} of X-ray mirror fabrication in the future, not only in terms of manufacture, but also in terms of design.        

\bibliography{report}
\bibliographystyle{spbasic}

\end{document}